\documentclass[preprint,10pt]{aastex}
\usepackage{emulateapj5}
\renewcommand{\*}{$^*$}
\newcommand{\Jb}{\mbox{Jy beam$^{-1}$}}
\newcommand{\uv}{\mbox{$u$-$v$}}
\newcommand{\ex}[1]{\mbox{$\times 10^{#1}$}}

\newcommand{\Msol}{\mbox{$M_\sun$}}
\newcommand{\Rsol}{\mbox{R$_\sun$}}
\newcommand{\kms}{\mbox{km s$^{-1}$}}
\newcommand{\muas}{\mbox{$\mu$as}}

\newcommand{\sbg}{\mbox{$\,\sigma_{\rm bg}$}}
\newcommand{\Ra}[4]{\mbox{${#1}^{\rm h} \; {#2}^{\rm m} \; {#3}\fs{#4} $}}
\newcommand{\dec}[4]{\mbox{${#1}\arcdeg \; {#2}\arcmin \; {#3}\farcs{#4} $}}
\newcommand{\thout}{\mbox{$\theta_{\rm o}$}}
\newcommand{\thin}{\mbox{$\theta_{\rm i}$}}
\newcommand{\Tb}{\mbox{$T_b$}}
\newcommand{\ve}{\mbox{$\epsilon_v$}}
\newcommand{\veunit}{\mbox{erg~cm$^{-3}$ Hz$^{-1}$ s$^{-1}$}}

\shortauthors{Bietenholz et al.}
\shorttitle{SN~1993J VLBI III}

\begin{document}

\title{SN 1993J VLBI (III): The Evolution of the Radio Shell}

\author{M. F. Bietenholz\altaffilmark{1}, N. Bartel\altaffilmark{1},
and M. P. Rupen\altaffilmark{2}
}

\altaffiltext{1}{Department of Physics and Astronomy, York University, Toronto,
M3J~1P3, Ontario, Canada}
\altaffiltext{2}{National Radio Astronomy Observatory, Socorro, New Mexico
87801, USA}

\begin{abstract}

A sequence of images of supernova 1993J at 30 epochs, from 50 d to
$\sim 9$~yr after shock breakout, shows the evolution of the expanding
radio shell of an exploded star in detail. The images were obtained
from 24 observing sessions at 8.4~GHz and 19 at 5.0~GHz and from our
last session at 1.7 GHz. The images are all phase-referenced to the
stable reference point of the core of the host galaxy M81.  This
allows us to display them relative to the supernova explosion
center. The earliest image shows an almost unresolved source with a
radius of 520~AU.  The shell structure becomes discernible 175~d after
shock breakout.  The brightness of the ridge of the projected shell is
not uniform, but rather varies by a factor of two, having a distinct
peak or maximum to the south-east and a gap or minimum to the west.
Over the next $\sim 350$~d, this pattern rotates counter-clockwise,
with the gap rotating from west to north-northeast.
After two years, the structure becomes more complex with hot spots
developing in the east, south, and west.  The pattern of modulation
continues to change, and after five years the hot spots are located to
the north-northwest, south and south-southeast.  After nine years,
the radio shell has expanded to a radius of 19,000~AU\@.
The brightness in the center of the images is lower than expected for
an optically thin, spherical shell.  Absorption in the center is
favored over a thinner shell in the back and/or front.  Allowing for
absorption, we find the thickness of the shell is $25 \pm 3$\% of its
outer radius.  We place a $3\sigma$ upper limit of 4.4\% on the mean
polarization of the bright part of the shell, consistent with internal
Faraday depolarization.  We find no compact source in the central
region above a brightness limit of 0.05~m\Jb\ at 8.4~GHz,
corresponding to 30\% of the current spectral luminosity of the Crab
Nebula.  We conclude that either any pulsar nebula in the center of
SN~1993J is much fainter than the Crab or that there is still
significant internal radio absorption.
\end{abstract}


\section{INTRODUCTION}

SN 1993J is one of the brightest radio supernovae ever detected (e.g.,
Weiler et al. 2002).  Its location in the nearby spiral galaxy M81
high in the northern sky has made it the best target for VLBI
supernova studies so far.  This paper is the third in a series
presenting the results from our VLBI campaign on this supernova
(Bietenholz, Bartel \& Rupen, 2001, Paper~I; Bartel et~al.\ 2002,
Paper~II), and for the convenience of the reader we repeat the
following two introductory paragraphs from Paper~II below.

\objectname[]{SN~1993J} was discovered in a spiral arm of
\objectname[]{M81} south south-west of the galaxy's center by Garcia
(Ripero \& Garcia 1993) on 28 March 1993 shortly after shock breakout
at $\sim 0$~UT ($t=0$) on the same day (Wheeler et~al.\ 1993). It
subsequently became the optically brightest supernova in the northern
hemisphere since SN~1954A.  At a Cepheid-distance of $3.63 \pm
0.31$~Mpc (Ferrarese et al.\ 2000; see also Freedman et~al.\ 1994,
2001),
it is also one of the closest extragalactic supernovae ever observed
and is second only to SN~1987A as a subject of intense observational
and theoretical supernova studies.  The precursor was identified soon
after the supernova discovery (e.g., Humphreys et~al.\ 1993) and found
to be an approximately K0~I supergiant with a likely mass of $\sim$ 17
\Msol\ (Aldering, Humphreys, \& Richmond 1994; Van Dyk et al.\ 2002)
and a radius $\geq$ 675 \Rsol\ or $\geq 3.5$~AU (Clocchiatti et~al.\
1995).

The light curve and the spectral properties indicated that SN~1993J
was of Type~IIb, characterized by a low-mass hydrogen outer
layer. H\"oflich, Langer, \& Duschinger (1993) considered a massive
($\sim 30$~\Msol) single supergiant that had lost most of its hydrogen
envelope through a strong wind and retained an envelope mass of $\sim
3$~\Msol\ (but see also H\"oflich 1995).  However it is more likely
that the progenitor had the lower mass given above, and a close binary
companion that stripped off most of its hydrogen envelope. The
progenitor was then left with a residual hydrogen mass in the outer
shell probably in the range of 0.2 to 0.4~\Msol\ (e.g., Podsiadlowski
et~al.\ 1993; Woosley et~al.\ 1994; Houck \& Fransson 1996) with some
estimates being larger, but none greater than 0.9 \Msol\ (Nomoto
et~al.\ 1993; Shigeyama et~al.\ 1994; Bartunov et~al.\ 1994).

The ejecta thrown off in the supernova explosion expand into, and
interact with, the circumstellar medium (CSM), which is expected to
consist of the slow, dense wind of the progenitor.  In the period
before the star died, the progenitor is thought to have had lost mass
with a mass-loss to wind-velocity ratio $\dot
M/w=5\ex{-5}$~\Msol/10~\kms\ (Van Dyk et~al.\ 1994; Fransson \&
Bj\"ornsson 1998).  As the ejecta expand into the CSM, it is expected
that a forward shock is driven into the surrounding CSM, and a reverse
shock driven back into the expanding ejecta.  The radio emission is
synchrotron emission, which is most likely generated by relativistic
particles accelerated in the region between these two shocks (e.g.,
Chevalier 1982a).

In the case of SN~1993J, there may have been asymmetries in the ejecta
and anisotropic expansion velocities. In particular, asymmetric
spectral lines were observed (e.g., Lewis et~al.\ 1994; Spyromilio
1994), and significant time-variable optical polarization was found in
the spectra (Trammell, Hines, \& Wheeler 1993; Tran et~al.\
1997). While the asymmetry in at least one line was possibly caused by
line blending (Houck \& Fransson 1996), the detection of the
polarization is a strong argument for asymmetry in the optical
emission region and led to ejecta models with non-spherical
geometries (e.g., H\"oflich 1995; H\"oflich et~al.\ 1996).

There are a number of different mechanisms which might produce
asymmetry in the ejecta of a supernova.  Some of them operate even
before the shock breaks out through the surface of the progenitor.
For instance, an axisymmetric density distribution of the progenitor
might lead to an asymmetric explosion and an anisotropic expansion
pattern.  Also, recent numerical modeling of massive star explosions
suggests that shortly after the bounce, the expansion has
fundamentally anisotropic components and develops ``fingers'' with
speeds twice the average expansion speed.  If the CSM is anisotropic,
as is expected of some red giant winds, further anisotropy could
develop after shock breakout.  Finally, the contact surface between
the ejecta and the CSM is subject to the Rayleigh-Taylor instability,
which, with time, could lead to fingers of shocked envelope material
extending into the shocked CSM\@.  A detailed sequence of images of
the radio structure of the supernova is therefore of particular
importance because it may allow us to see some of these mechanisms in
action.

Early VLBI observations of SN~1993J allowed the size of the partly
resolved supernova to be determined (Bartel et~al.\ 1993, 1994;
Marcaide et~al.\ 1993, 1994). The radio source was shown to be circular
within 5\% (Bartel et~al.\ 1994). As the supernova expanded further, a
shell morphology could be discerned (Marcaide et~al.\ 1995a, b;
Bartel, Bietenholz, \& Rupen 1995).  More accurate measurements over
seven years of observations allowed us to show that the radio shell's
20\% contour was circularly symmetric even within 3\%, and that its
angular expansion from the explosion center isotropic within 5.5\%
(Paper I).  The observed highly isotropic expansion from the explosion
center is in stark contrast to the anisotropic expansion suggested by
the optical observations.  In this context, a detailed investigation
of the structure of the radio shell and its evolution with time
becomes of particular importance.

The first angular expansion velocity determinations, when compared
with optical velocity measurements, showed that the radio emission
emanates from the shock region (Bartel et~al.\ 1994).
Subsequently, the expansion of the supernova underwent several
changes.  At 30 days after shock breakout ($t = 30$~d), the shell
was expanding at $\sim 17,200$~\kms.  From then till $t \sim 300$~d,
the shell was slightly decelerated, with the outer radius, $\thout
\propto t^m$ and $m = 0.919 \pm 0.019$.  Then the deceleration grew
significantly till $t \sim 1600$~d, with $m$ decreasing to $\sim 0.74$
and the expansion velocity slowing to $\sim 8900$~\kms.  Subsequently,
the deceleration lessened again, with $m$ increasing to $\sim 0.85$.
These changes were related to changes of the radio light-curves and
the spectra, and interpreted in terms of a stratification of the
ejecta, with the high-mass ejecta starting to pass through the reverse
shock and exerting greater pressure on the shocked low-mass envelope
and the shocked CSM (Paper II; see also Mioduszewski, Dwarkadas, \&
Ball 2001).

In Paper~I, we located the explosion center with respect to the core
of the core-jet source M81\*, thus defining a stable reference point
for our images.  We also determined, using model-fitting, an upper
limit to the proper motion of the geometric center of SN~1993J\@,
and consequently on anisotropic expansion of the radio shell.
In Paper~II, we determined the expansion speed of SN~1993J, measured
its deceleration, and studied the radio light curves and the related
changes in the radio spectrum.  In this third paper, we present a
complete series of VLBI images of SN~1993J at 8.4 and 5.0~GHz, along
with our latest image at 1.7~GHz.  While some of these images have
already been presented earlier (see Papers I, II; Bietenholz et~al.\
2001; Bartel et~al.\ 2000; Bartel, Bietenholz, \& Rupen 1995), we
present here a complete and uniform set of images from 50~d after
shock breakout till the present.  These images form the most complete
set of images of an expanding supernova ever obtained.  (For some
parallel observations with up to seven consecutive images, see Marcaide
et~al.\ 1995b, 1997, 2002).

\section {OBSERVATIONS AND DATA REDUCTION\label{sobsdat}}

From 1993 to 2001 we made 66 multi-frequency VLBI observations at 34
epochs at 22.2, 15.0, 8.4 , 5.0, 2.3, and 1.7 GHz. We used a global
array of between 11 and 18 telescopes with a total time of 12 to 18
hours for each epoch. These observations were described in Papers I
and II\@.  At 22.2 and 15.0 GHz, the supernova was only bright enough
for useful observations in the first year.  Even at the higher
resolution available in principle at these frequencies, only the size
but not yet the structure of the source could be determined.  In
particular, the gain in angular resolution at 15~GHz over that at
8.4~GHz with a comparable array was offset by the unavailability of
European VLBI Network (EVN) or Deep Space Network (DSN) antennas at
the higher frequency.  In consequence, even during the first year, we
obtained the best data for imaging at 8.4~GHz.

\begin{figure}
\plotone{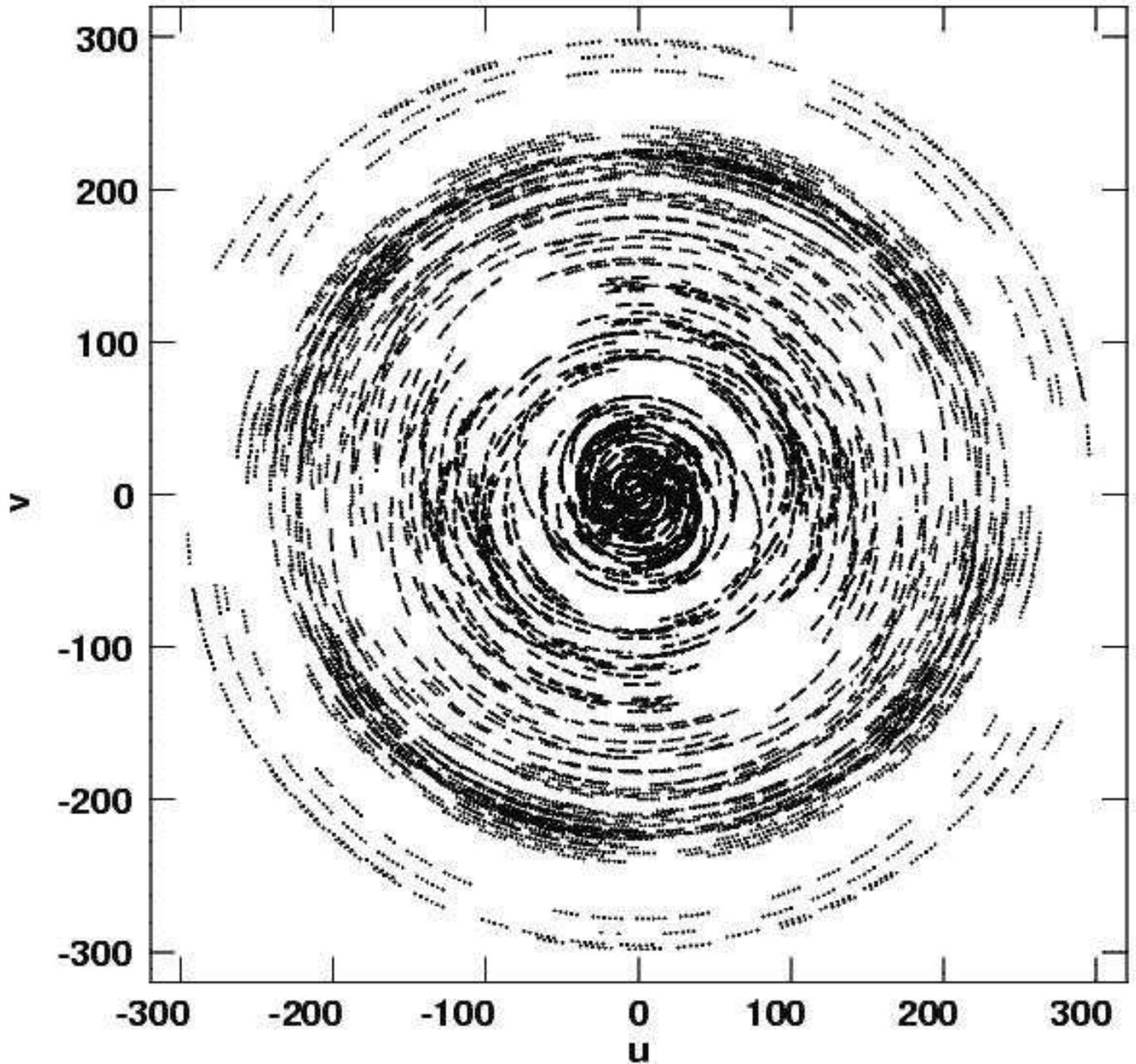}
\figcaption{The \uv~coverage of our observations at $t = 2787$~d (2000
November 13) at 8.4~GHz, showing the dense and nearly circular
\uv~coverage.  Both $u$ and $v$ are given in Mega-wavelengths.
\label{fuvcov}}
\end{figure}

The high declination of 69\arcdeg\ of SN~1993J enabled us to obtain
essentially 100\% visibility at almost every telescope, and as a
result dense, fairly uniform, and nearly circular \uv~coverage for
many of our observations.  In Figure~\ref{fuvcov} we show as an
example the \uv~coverage for the 8.4~GHz observations on 2000 November
13 ($t = 2787$~d).  Most of the observations were made by
phase-referencing to M81\* (Bietenholz, Bartel, \& Rupen, 2000),
providing a combination of unsurpassed sensitivity for imaging and
accurate astrometry for tracking the position of the explosion center
in the images for essentially all epochs.

We used a global array of between 9 and 18 telescopes with a total
time of 9 to 18 hours for each run\footnote{see Paper~II for further
details of the arrays used at each session except for the 5~GHz epoch
at 2002 May 25, for which the details are identical to that of 1999
June 16, with the exception that Nt did not observe, and that the
total time was 11.8 hours and the on-source time was 439
baseline-hours.}.
The data were recorded with either the MK~III or the VLBA/MKIV VLBI
systems, and correlated with the NRAO VLBA processor in Socorro, New
Mexico, USA.  The analysis was carried out using NRAO's Astronomical
Image Processing System (AIPS).  The usual procedure in calibrating
VLBI data is to determine the instrumental phases, which are
essentially unknown, by self-calibration using an arbitrary starting
model which is typically a point source (Walker 1999).  While this
procedure generally converges well, it can introduce symmetrizing
artifacts into the images (Linfield 1986; Massi \& Aaron 1999).
Phase-referencing allowed us to calibrate the instrumental phases with
respect to M81\* to the extent that we no longer needed to
self-calibrate with an arbitrary starting model, and thus avoided any
symmetrization.  We are thus assured of the most un-biased images
possible.  In those cases where the signal-to-noise ratio was
sufficiently high we proceeded to self-calibrate in phase, using the
phase-referenced image as a model, to further improve the images.  In
fact, for later epochs phase-referencing was necessary for any imaging
of SN~1993J, since the decreasing flux density made self-calibration
impossible.

All the images were deconvolved with the CLEAN algorithm, using the
robust weighting scheme implemented in the AIPS task IMAGR (Briggs
1995; Briggs, Schwab, \& Sramek 1998).  For imaging, the (calibrated)
complex correlation coefficients are usually weighted by the inverse
of the thermal noise variance.  Since residual calibration errors may
be present in the data and need not scale with the thermal noise, we
compressed the weights somewhat by weighting with the inverse of the
rms, rather than the usual variance, of the thermal noise.

Despite the mostly superb \uv~coverage, differences from epoch to
epoch in the number of antennas and their scheduled time for the
observations caused variations in the effective angular resolution at
each frequency.  In particular, the parameters of an elliptical
Gaussian fit to the inner portion of the ``dirty beam'' varied from
epoch to epoch.  We aimed for the most consistent representation of
the images in each sequence to facilitate inter-comparison of the
images, and to minimize any possible misinterpretation due to a
varying angular resolution.  For this purpose, we present the
sequences of images at 8.4 and at 5.0~GHz each convolved with circular
Gaussian restoring beams whose widths increase monotonically with
time.  Since the resolution is naturally somewhat lower at 5.0~GHz
than at 8.4~GHz, a somewhat larger beam was used at 5.0~GHz in the
early images.  In the later images, the increased dynamic range at
5~GHz compared to that at 8.4~GHz allowed us to use the same restoring
beam at both frequencies, which allows for better comparison of the
images at the two frequencies.  In general, we chose the width of the
convolving beam to be approximately equal to, or somewhat larger than,
the maximum axis of the elliptical Gaussian fit to the inner portion
of the ``dirty beam.''  In a few early cases, we mildly super-resolved
our images along one or both axes of the dirty beam, and these cases
are noted in Table~\ref{tsnmaps}.  In the later images, because of the
low signal-to-noise ratio, we choose a restoring beam somewhat larger
than the inner portion of the dirty beam.  In the remainder of this
paper, when we refer to the resolution of an image, we mean the full
width at half maximum (FWHM) of the Gaussian convolving
beam\footnote{The exact procedure we used to make our final images was
as follows: We first adjusted the weighting scheme via the Briggs'
robustness parameter to achieve a dirty beam with FWHM close to but
somewhat smaller than the desired resolution.  We then followed the
usual procedure in making a restored CLEAN image of convolving the
CLEAN components with the elliptical Gaussian ``CLEAN beam'' fitted to
the inner portion of the dirty beam, and then adding the
(un-deconvolved) residuals from the CLEAN deconvolution.  We finally
re-convolved the whole restored image, now including the residuals, to
the desired effective circular Gaussian resolution.  This procedure
serves to keep the residuals at approximately the same effective
resolution as the CLEAN components.  However, the CLEANing process was
carried on well into the noise in all cases, so any contribution from
un-CLEANed residuals over the extent of SN~1993J was minimal.}.

\section{THE SEQUENCES OF IMAGES \label{simages}}

\begin{figure}
\plotone{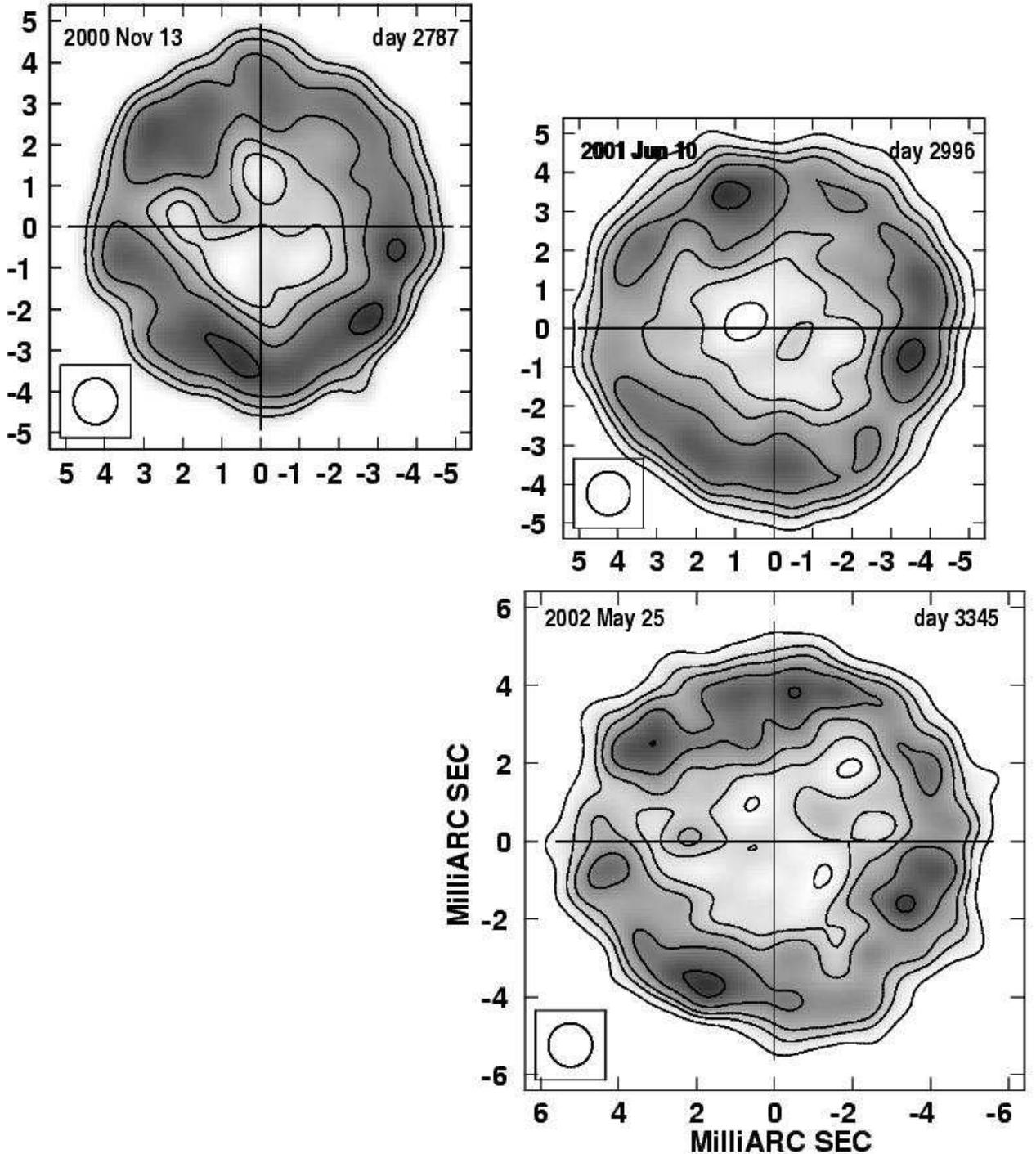}
\figcaption{VLBI images of SN~1993J at 8.4~GHz, in the left column,
and at 5.0~GHz, in the right column.  The resolution is shown at the
lower left in each panel.  The greyscale runs from 8 to 100\% of the
peak brightness for images for $t \leq 1253$~d, and from 16 to 100\%
of the peak brightness thereafter.  The contours are drawn at 1, 2, 4,
\dots, 32, 45, 64 and 90\% of the peak brightness, starting at the
first contour greater than $3 \, \sbg$, where \sbg\ is the standard
deviation of the background brightness.  See Table~\ref{tsnmaps} for
\sbg, the peak brightness, and the FWHM of the convolving beam.  At
most epochs, we switched between 8.4 and 5~GHz during the observing
run, and the images at each frequency are presented side by side with
the date and day number indicated only in the 8.4~GHz panel.  For the
remaining epochs, there were separate observing runs for 8.4 and
5~GHz, and these cases are represented by offset image panels. (See
Fig.~\ref{fcolor} for a false-color version of the 8.4~GHz images).
On this and on all our subsequent images, north is up and east to the
left, and the origin of the coordinate system, indicated by the large
crosses, is at the center of the fitted supernova shell.  On average,
the fitted centers are within 64~\muas\ rms of the explosion center at
8.4~GHz.  See Paper~I for the offsets of the fitted center and
explosion center for the individual images.
\label{fsnmaps}}
\end{figure}

\begin{figure}
\plotone{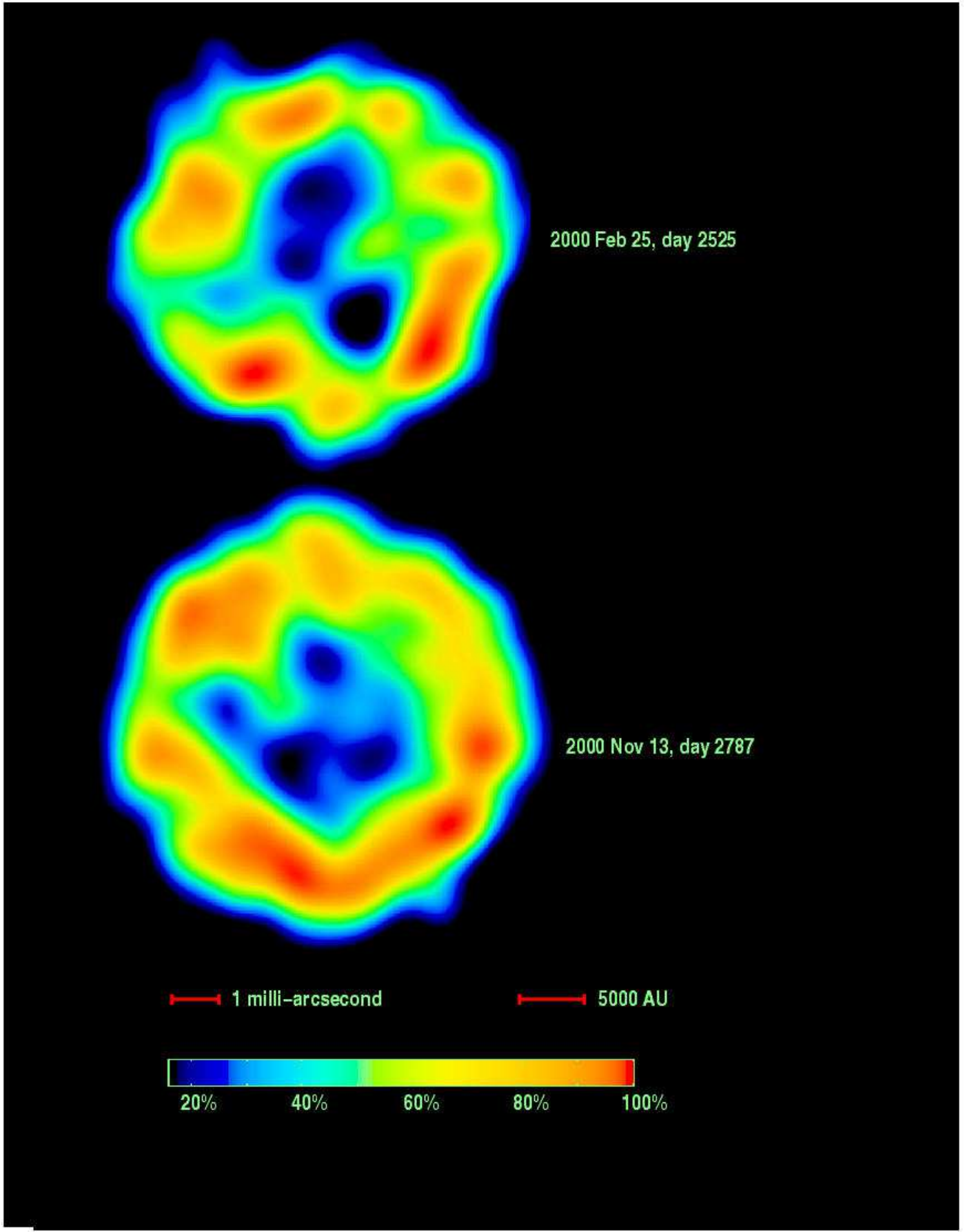}
\figcaption{The VLBI images of SN~1993 at 8.4~GHz in false colour.
The brightness scale, indicated at the end, runs from 16 to 100\% of
the peak brightness in each image.  Table~\ref{tsnmaps} lists the peak
brightness and the FWHM of the convolving beam for each image.
\label{fcolor}}
\end{figure}

\begin{figure}
\plotone{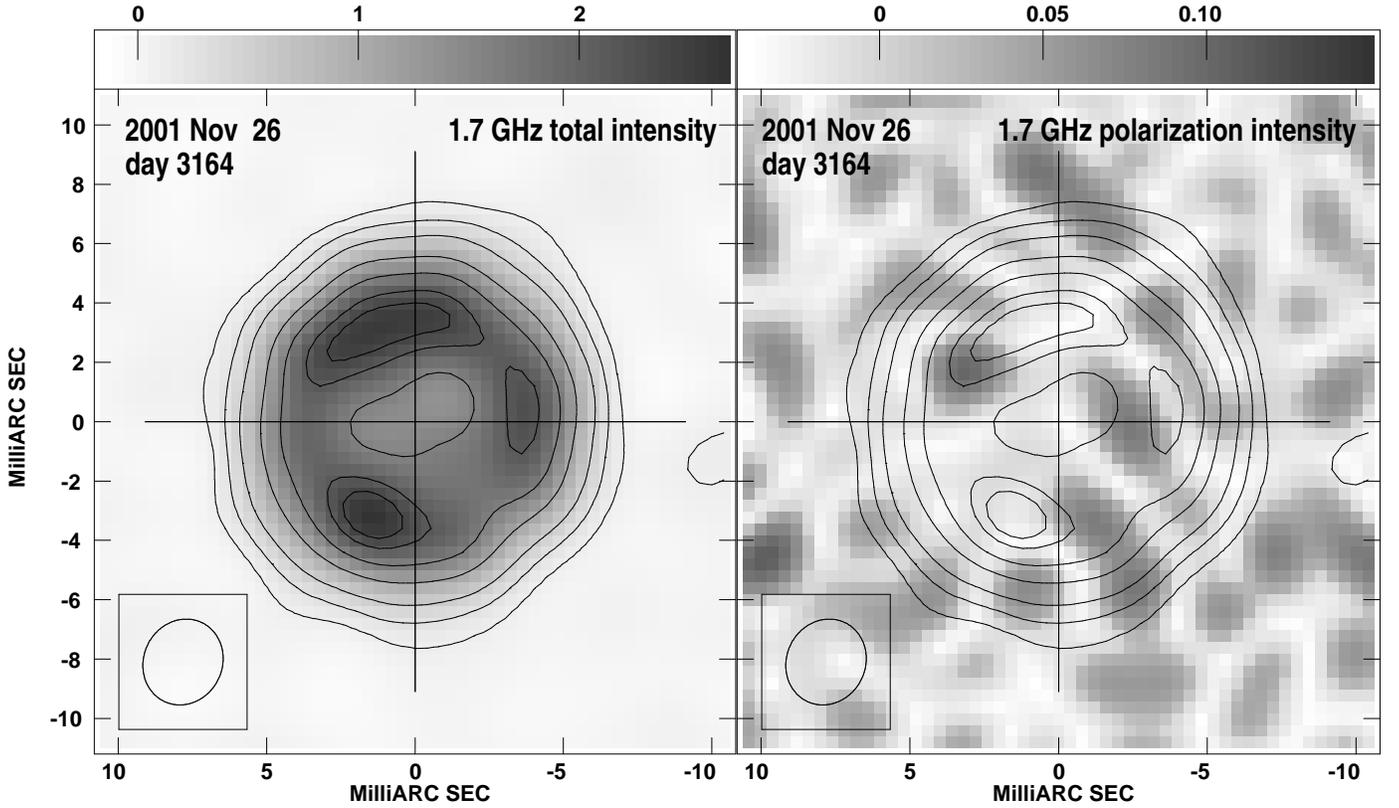}
\figcaption{Our latest 1.7~GHz VLBI image of SN~1993J, observed at $t
= 3164$~d.  The resolution is shown at the lower left and the
greyscales are labeled in m\Jb\ in both panels.
At left, we show the total intensity (Stokes $I$), with contours drawn
at $-3.6$, 3.6, 10, 20, 40, 60, 80, and 90\% of the peak brightness of
2.66~m\Jb.  The rms background brightness, \sbg, was 32~$\mu\Jb$.  At
right, we show the linear polarization intensity ($S_{\rm pol} =
\sqrt{Q^2 + U^2}$), corrected for the noise bias, and hence being
sometimes unphysically negative.  The total intensity contours are
repeated for reference.
\label{flimage}}
\end{figure}

In Figure~\ref{fsnmaps} we display 24 images of the supernova at
8.4~GHz along with 19 images at 5.0~GHz.  In addition, we display the
8.4~GHz images in false color in Figure~\ref{fcolor}.  In
Table~\ref{tsnmaps} we give key characteristics of each of these
images.  In Figure~\ref{flimage} we show our most recent image,
observed at 1.7~GHz on 2001 November 26 ($t = 3164$~d).  In each
image, we take the geometric center of the supernova shell as the
origin of the coordinate system.  In particular, we describe in
Paper~I how we fit a geometric spherical shell model\footnote{The
model consisted of the two-dimensional projection of a
three-dimensional spherical shell with uniform volume emissivity and
with the ratio of the outer to inner angular radius, $\thout / \thin =
1.25$ corresponding to a shell thickness, $\thout - \thin$, of $0.2 \,
\thout$ (see Paper~I for details). The use of a circular model despite
the evident modulation of the shell brightness is justified because,
as we show in Paper~I, SN~1993J remains circular to within 3\%.
Furthermore, as we show in Paper~II, the fit size of the spherical
shell evolves very smoothly, which suggests that, despite the
modulation of the brightness around the ridge of the projected shell,
the uniform spherical shell model is a good overall description of
SN~1993J's radio emission.}
directly to the \uv~data.  It is the center of this fit model we
take as the geometric center of the shell.

In Paper I, we determined the coordinates of the explosion center.  At
8.4~GHz, the positions of the geometric center at each epoch are
tabulated as offsets from the coordinates of the center of explosion,
$\alpha_{\rm explosion} = \Ra{09}{55}{24}{7747593}, \; \delta_{\rm
explosion} = \dec{69}{01}{13}{703188}$ (J2000), for all but the last
two epochs for which the data reduction was not complete at the time
that Paper I was published.  The rms variation of these offsets is
64~\muas, and is
not significantly different from the combined standard errors of the
shell center and explosion center positions of 60~\muas.

The images show the dynamic evolution of the expanding radio shell
from 50~d to 3345~d after the explosion.  Never before has it been
possible to obtain such detailed information on the radio emission
from a supernova.  With phase-referencing, dense \uv~coverage, and the
use of essentially the most sensitive array available for such
observations, the images are of the highest quality that could be
obtained for the time of the observations. The most sensitive
observations were those of our last epoch at 5.0~GHz. The standard
deviation of the background brightness was just 16~$\mu\Jb$ and the
peak brightness in the image was 420~$\mu\Jb$ (see
Table~\ref{tsnmaps}), giving us a nominal dynamic range of 26, among
the highest ever obtained for such a weak source.

\begin{figure}
\plotone{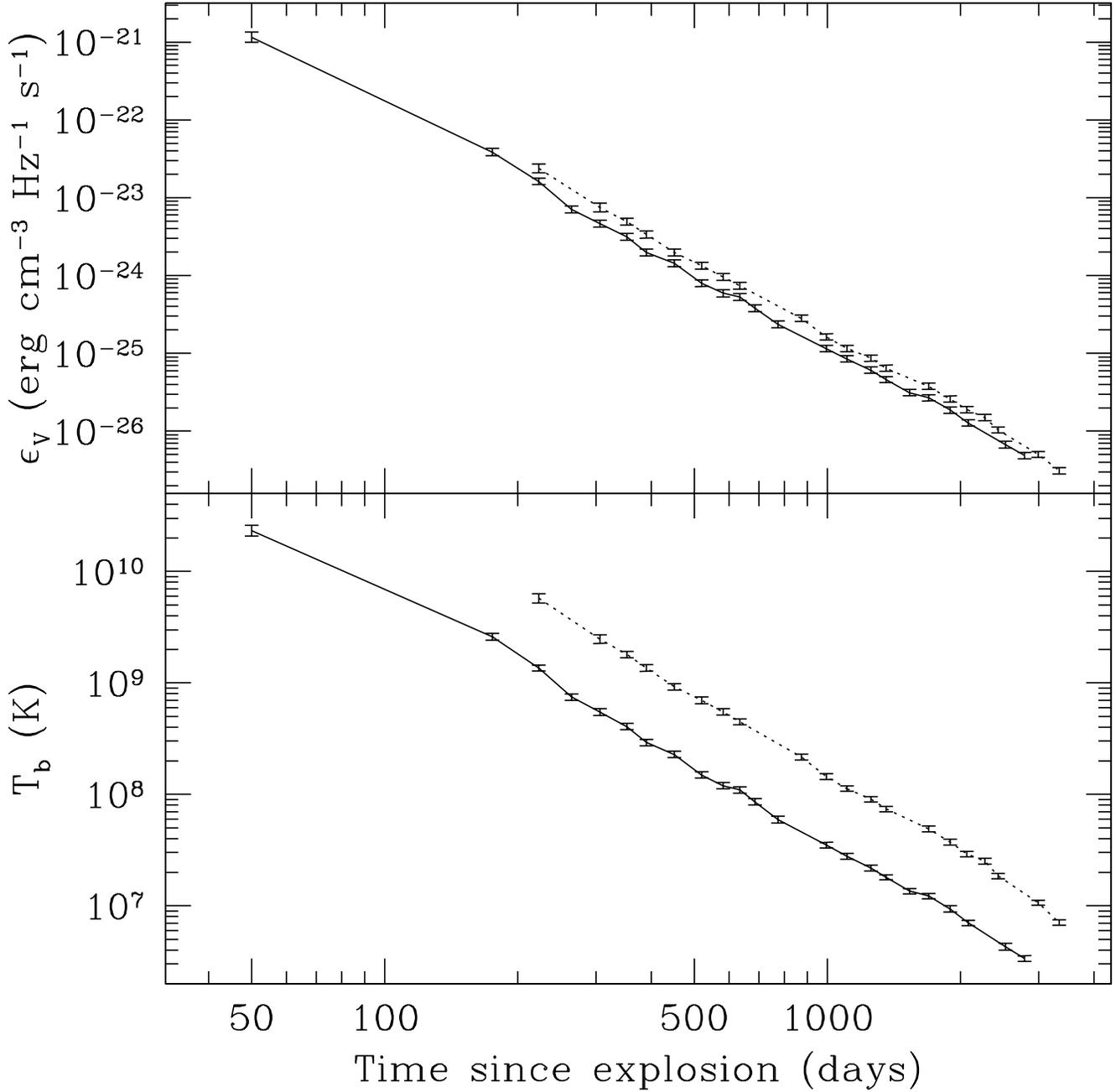}
\figcaption{The mean spectral volume emissivity, \ve, and brightness
temperature, \Tb, of SN~1993J as a function of time at 8.4 and
5.0~GHz. We calculate \ve\ and \Tb\ from the total flux densities
measured at the VLA by taking the outer angular radius of SN~1993J to
be the fit value of \thout\ (see Paper II).  For \ve, we take a
distance of 3.6~Mpc.  We also assume a spherical shell with a ratio of
the outer to the inner radius of 1.34 and without any absorption.  The
uncertainties are approximately standard errors.
\label{ftb}}
\end{figure}

In Figure~\ref{ftb} we plot the mean spectral volume emissivity, \ve,
and brightness temperature, \Tb, for SN~1993J for each of the images
in Fig.~\ref{fsnmaps}.  We calculate \ve\ and \Tb\ from the total flux
densities and fit outer radii, \thout, from Paper~II\@.  For \ve\ we
take a distance of 3.6~Mpc and assume a spherical shell with ratio of
the outer to the inner radius of 1.34.
At 8.4~GHz, \ve\ declines from a value of $(1.1 \pm 2)$\ex{-22}
\veunit\ at $t = 50$~d to $(4.9 \pm 0.5)$\ex{-27} \veunit\ at $t =
2787$~d.  The behavior at 5.0~GHz the is very similar, but the values
are $\sim 40$\% higher.  The behavior of \Tb\ is very similar to that
of \ve, and at 8.4~GHz \Tb\ declines from a value of $(2.3 \pm
0.3)$\ex{10}~K at $t = 50$~d to $(3.3 \pm 0.2)$\ex{6}~K at $t =
2787$~d.  The peak brightness temperatures, assuming features the size
of the beam, are higher than the average values by a factor of $\sim
2$ for the epochs at which SN~1993J was clearly resolved.

\subsection{General Aspects of the Changing Brightness Distribution
\label{saspects}}

\begin{figure}
\plotone{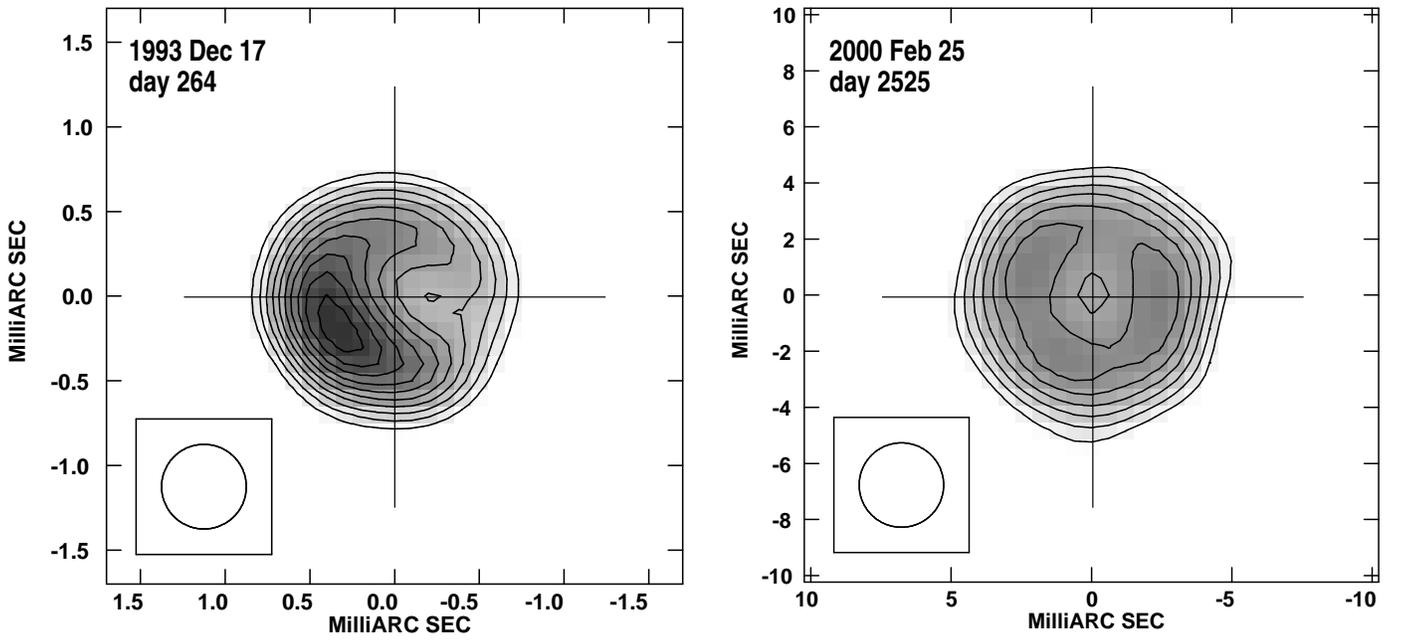}
\figcaption{A comparison of two 8.4-GHz images of SN~1993J convolved
to the same resolution relative to the outer angular radius of the fit
shell.  Left: the image from $t = 264$~d, convolved to 0.5~mas
resolution.  Right: the image from $t = 2525$~d, convolved to 3.0~mas
resolution.  The contours are comparable in both images and are drawn
at 6, 8, 10, \dots\ 24\% of the total flux density per
beam area.  The greyscale is labeled in m\Jb\ and runs from 5 to 25\%
of the total flux density per beam area. 
\label{fdec93-feb00}}
\end{figure}

Before we focus on the individual images and specific features
therein, we study the general aspects of the evolving brightness
distribution of the expanding radio shell.  At early epochs, with $t <
175$~d, SN~1993J is still unresolved. The shell structure first
becomes visible at $t = 175$~d, and it remains so for all our
subsequent images.  As we have already shown in Papers~I and II, the
shell remains remarkably circular in projection.  Perhaps the most
striking overall feature of the structure is that the brightness of
the projected shell is quite strongly modulated in position angle,
p.a., despite the circularity of the outer contours.  In fact, even at
the earliest epoch where the shell is resolved ($t = 175$~d) the
brightness around the ridge varies by a factor of $\sim 2$, with a
pronounced maximum or peak at p.a.\ $\sim 135$\arcdeg\ and a minimum
or gap at p.a.\ $\sim -90$\arcdeg.  At late times, the modulation has
become more complex, and there is no longer such a clear one-sided
pattern.  To illustrate this behavior more clearly, we show in
Figure~\ref{fdec93-feb00} the 8.4~GHz images at $t = 264$ and 2525~d,
both convolved to the same relative resolution of $0.73 \times$ the
shell outer angular radius as determined in Paper~II\@. The shell
appears much more uniform at $t = 2525$~d.

We defer detailed discussion of the reliability of the images and the
uncertainties to the following sections, but we make some general
remarks here.  A strong argument for the reality of the observed
modulation of the shell brightness is the excellent
correspondence between the images at 8.4 and 5.0~GHz which is clearly
visible in Figure~\ref{fsnmaps}.  The slow changes from
epoch to epoch visible also in Figure~\ref{fsnmaps} further argue for
the reality of the structure observed.

\begin{figure}
\plotone{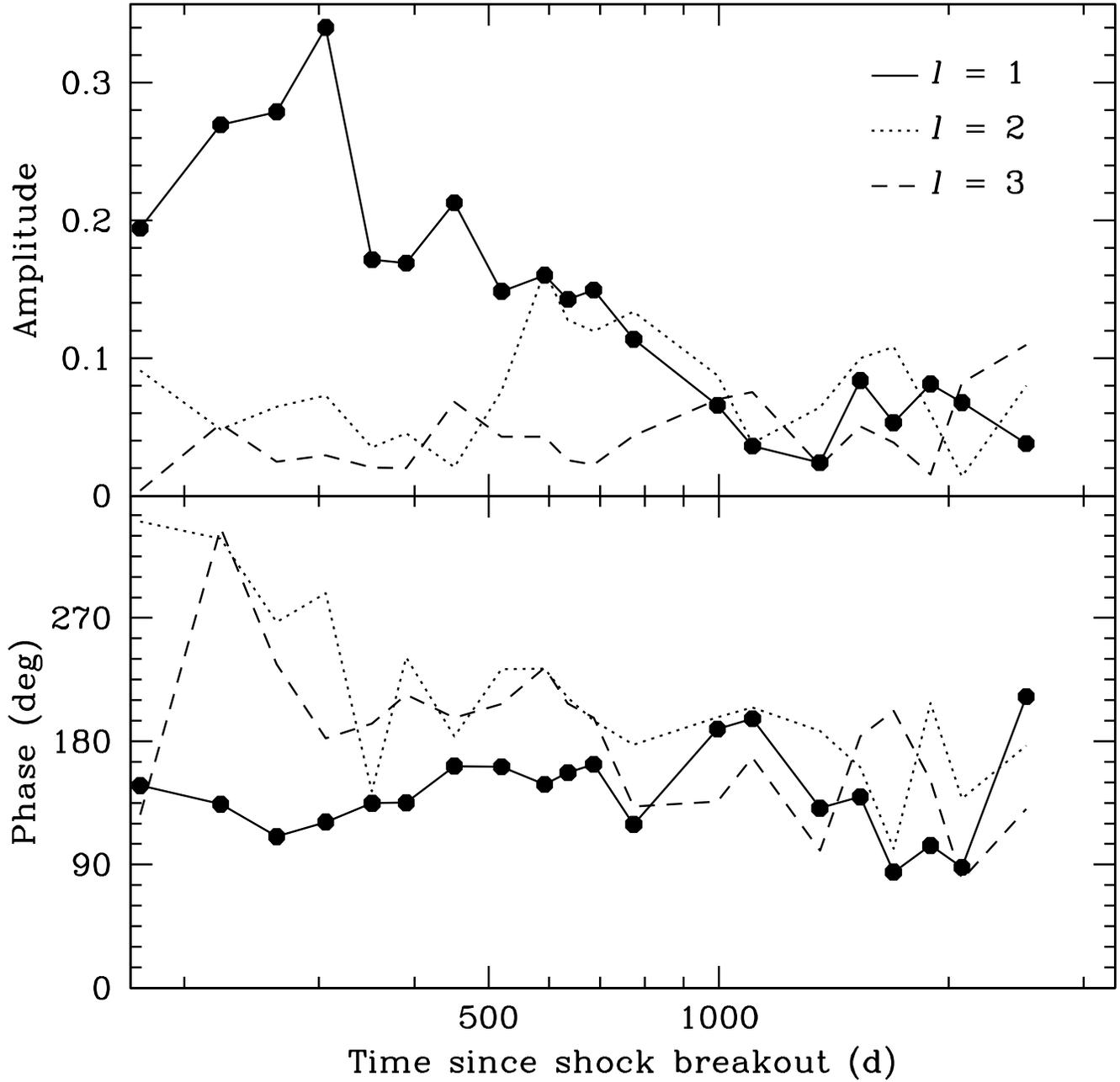} 
\figcaption{The amplitudes and phases of the first three harmonics ($l
= 1,2,3$) of the Fourier expansion of the modulation of the brightness
of the radio shell at 8.4~GHz as a function of position angle, p.a.,
(integrated in radius).  The wave number, $l$, is the number of maxima
along the circumference. For $l = 1$, the phase is the p.a.\ of the
maximum.  For $l > 1$, the phase is the p.a.\ of the first maximum
multiplied by $l$. \label{fl13mod}}
\end{figure}

In order to discuss the modulation of the brightness around the ridge,
we introduce the following simple parameterization.  For simplicity,
we consider only modulation of the two-dimensional brightness
distribution with p.a.  Let $l$ be the wave number of a sinusoidal
modulation with p.a.\ of the projected shell, such that there are $l$
maxima around the circumference.  We plot the amplitude and phase of
the first three sinusoids with $l = 1, 2, 3$ at 8.4~GHz, at which
frequency the resolution is higher for early epochs, in
Figure~\ref{fl13mod}.

At $t = 264$~d the amplitude of the $l = 1$ modulation is quite
large, and increases further till $t = 306$~d.  It is possible that
this early rise is a resolution effect, since a finite resolution will
tend to suppress the apparent modulation.  After $t = 306$~d, the
$l=1$ amplitude falls steadily till $t \sim 1500$~d, which cannot
be ascribed to a resolution effect, but rather reflects the decreasing
asymmetry in brightness illustrated in Fig.~\ref{fdec93-feb00}.
The $l = 2$ amplitude shows a sharp rise at $t \sim 500$~d which
reflects the development of eastern and western hot spots discussed
further in \S~\ref{stwospots} below.  At late times the amplitudes of
$l = 1$, 2 and 3 are comparable, reflecting the increased complexity
of the images.

\subsection{Uncertainty in the Images \label{suncert}}

In order to discuss the images in detail, we first elaborate on the
brightness uncertainty in them.  The simplest estimator of this
uncertainty is the standard deviation of the brightness in empty
regions of the image, which we will call \sbg, and which we list for
each image in Table~\ref{tsnmaps}.  However, for the reasons detailed
below, \sbg\ is likely somewhat of an underestimate of the true
brightness uncertainty (see also Perley 1999).

The true uncertainty will have three principal components: 1) the
effect of the thermal noise, 2) the effect of residual calibration
errors, and 3) the effect of instabilities or inaccuracies in the
deconvolution process, which could also be described as the effect of
incomplete \uv~coverage, since in the case of complete \uv~coverage,
no deconvolution is necessary and therefore no such instabilities
or inaccuracies will occur.

We will discuss each of these components in turn.  The effect of the
thermal noise, 1), will be random, will be uniform over the image,
and will scale with the number and size of the telescopes, the
bandwidth, and the observing time.  It can be well estimated by \sbg.
Unfortunately, the effects of 2) and 3) are less predictable, and may
correlate with the actual structure in the image.  There is thus the
possibility that the effective uncertainty in our images over the
extent of SN~1993J is higher than \sbg.

The effect of residual calibration errors, 2), will be approximately
proportional to the total flux density.  We can estimate this effect
as follows.  Let $N$ be the number of antennas, $\sigma_\phi$ be the
rms of the residual phase mis-calibration in radians, and $M$ be the
number of independent time intervals.  We will conservatively take $M$
to be the number of hours of observing time to allow for correlated,
slowly varying calibration errors. In our phase-referenced, and in
some cases additionally phase self-calibrated data, the residual
calibration phase errors, $\sigma_\phi$ are almost certainly $<
0.5$~rad.  We do not separately calculate the effect of residual
amplitude calibration errors, since they are almost certainly smaller.
Perley (1999) gives an estimate of the dynamic range limit due to
residual mis-calibration of $\sqrt{M} N/ \sigma_\phi$, where the
dynamic range is the image peak brightness divided by \sbg.  For our
data, with $M \gtrsim 10$ and $N \gtrsim 11$, this gives a dynamic
range of $\gtrsim 60$.  This is probably a lower limit, since our
phase-referencing cycle time was much shorter than 1 hour, and thus a
realistic value of $M$ is likely larger than 10.  Since the observed
dynamic range is smaller than this conservative limit for all
except the earliest epochs, for which $\sigma_\phi$ is almost
certainly less than 0.5~rad due to the high signal-to-noise ratio and
accurate phase self-calibration, we can conclude that the uncertainty
introduced by residual calibration errors is likely small.

The effect of 3), instabilities or inaccuracies in the deconvolution
process, will be largely confined to the area over which CLEAN
components were sought in the deconvolution process, in other words
over the CLEAN window.  The effect will
scale with the completeness of the \uv~coverage.  We estimated the
effect of the instabilities or inaccuracies in the deconvolution as
follows: for a typical \uv~data set, the actual, measured,
visibilities were replaced by the Fourier transform of the SN~1993J
image.  A realistic amount of random noise was added to this
artificial data set, which was then deconvolved.  This process was
repeated for many realizations of the random noise.  We then computed
the rms variation over the different noise realizations at various
locations of the resulting images.  For the 8.4~GHz data sets at $t =
264$~d, 451~d, 1693~d and 2525~d, the standard deviation of the
brightness within the CLEAN window was higher by 20\%, 33\%, 16\%, and
37\% respectively, than that of the background.

We thus estimate the total image uncertainty in the CLEAN window, i.e.,
over the radio shell, to be between 1.2 and 1.4 times \sbg, or the
standard deviation of background brightness, the values of which are
listed in Table~\ref{tsnmaps}.

\subsection{Early Structure in the Shell and Apparent Rotation between 175 and 686 Days \label{searly}}

\begin{figure}
\epsscale{0.667}
\plotone{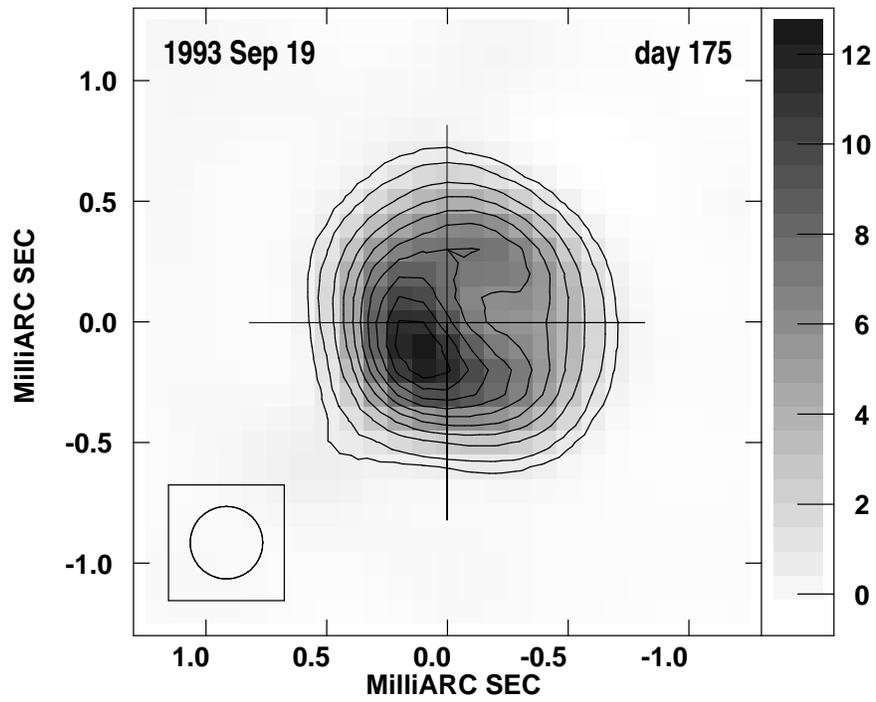} 
\figcaption{A high-resolution image of SN~1993J at 8.4~GHz at $t =
175$~d.  The resolution was 0.3~mas and is shown in the lower left.
The greyscale at right is labeled in m\Jb.  The contours are drawn at
$-5$, 5, 10, 20, 30, 40, 50, 60, 70, 80, and 90\% of the peak
brightness of 12.5~m\Jb.  Maximum entropy deconvolution was used, and
\sbg\ was 78~$\mu\Jb$.
\label{fsuperr}}
\end{figure}

The earliest 8.4~GHz image of SN 1993J, at $t = 50$~d, shows an almost
unresolved source with a radius of just 0.14~mas or 520~AU, only 100
to 200 times larger than the radius of the progenitor star.  No
brightness structure can yet be seen.  As mentioned above, the shell
structure first becomes visible at $t = 175$~d.  Already at this epoch
there is a distinct asymmetry in the brightness, with a maximum to the
east-southeast of the center and a minimum to the west.  To make the
asymmetry more visible, we display in Figure~\ref{fsuperr} the image
at this epoch with a moderate degree of super-resolution, achieved
using maximum entropy deconvolution as implemented in the AIPS task
VTESS, which generates less spurious structure than CLEAN (Briggs
1995).

\begin{figure}
\epsscale{0.75}
\plotone{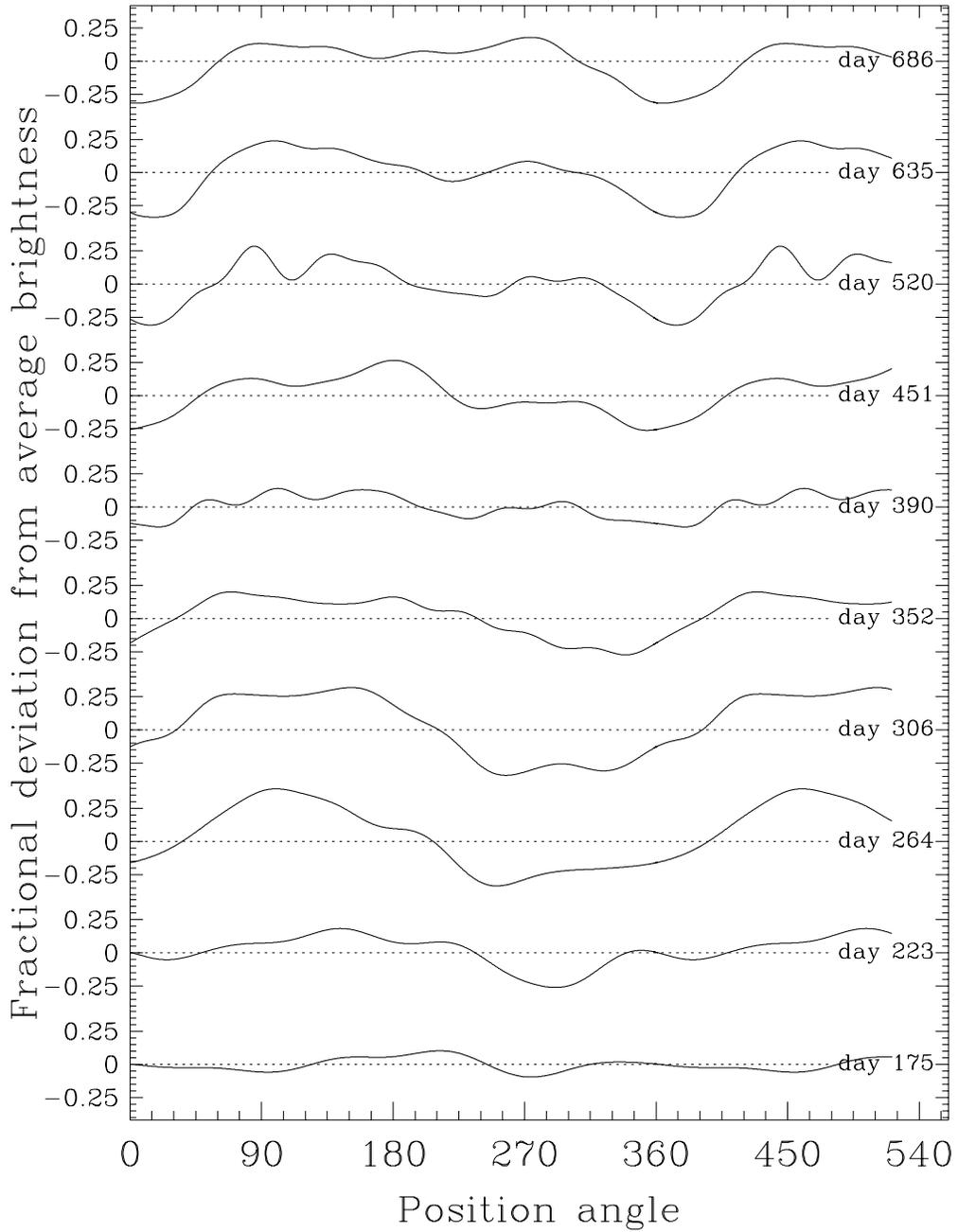}
\figcaption{The variation of the brightness along the ridge as a
function of position angle for several epochs of our observations,
showing the coherent evolution at early epochs.  Each of the curves
shows the fractional deviation from the average brightness, integrated
from 0.7 to $1.0\times$ the outer radius of the supernova as
determined by model fits.  The images used were those shown in
Fig.~\ref{fsnmaps}.  For a convenient display of the rotation of both
the maximum and the minimum, we extend the position angle axis to
540\arcdeg, causing some features to appear twice on each
curve.\label{fflxvpa}}
\end{figure}

\begin{figure}
\epsscale{1.0} 
\plotone{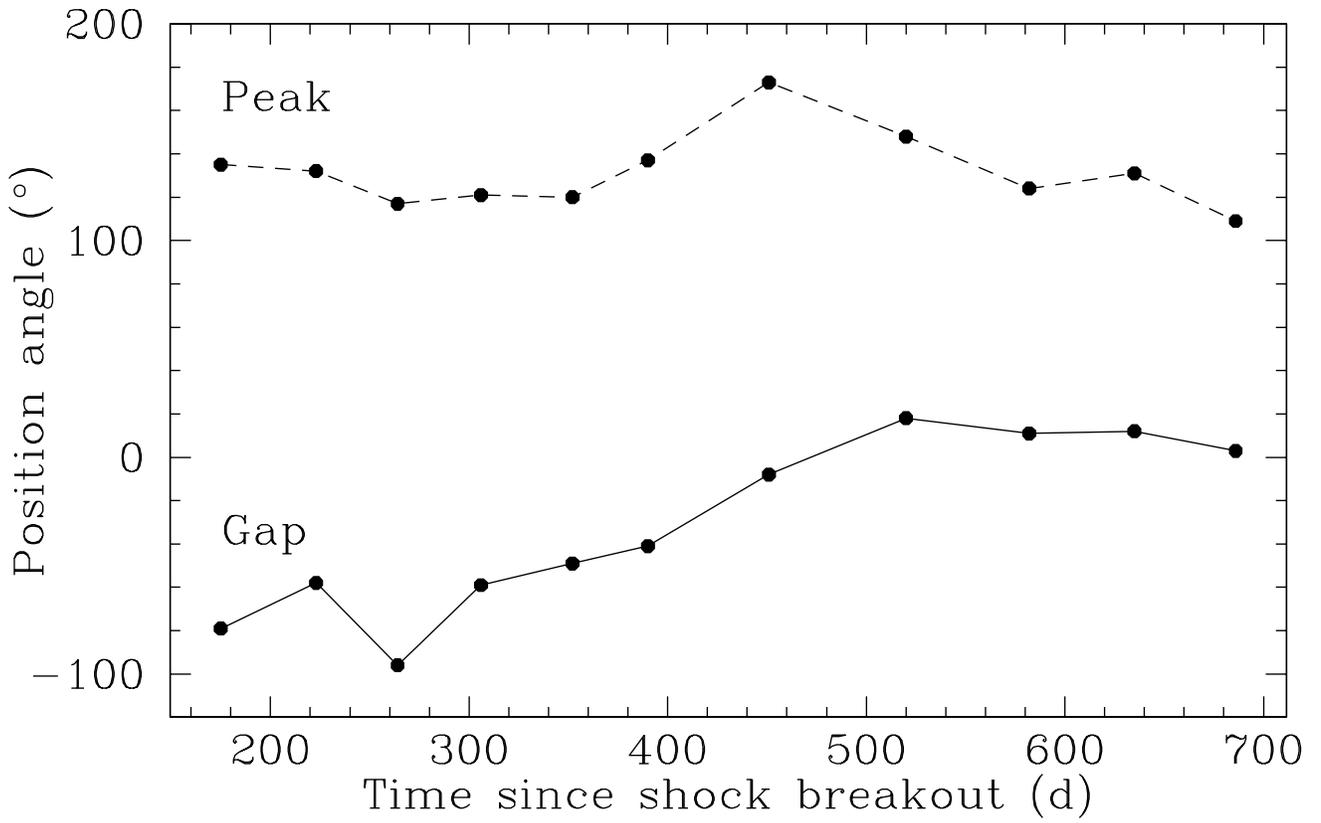} 
\figcaption{The position angle (p.a.) of the peak and gap in the
brightness of the ridge as a function of time.  The locations of the
peak and the gap were determined from the images in
Fig.~\ref{fsnmaps}, and the p.a.'s are referred to the geometric
center of the radio shell.  The p.a.\ of the gap was taken as that of
the midpoint of the ``horns'' of the 64\% contour, with the horns
being the points on the contours with the largest curvature.  The
uncertainties are estimated to be $\sim \pm 10$\arcdeg.
\label{fminmaxpa}}
\end{figure}

As the supernova expands, our relative resolution increases and the
pattern of modulation of the ridge brightness becomes clearer.  The
pattern also appears to rotate counter-clockwise.  By $t = 520$~d, the
brightness peak is to the south and the gap to the north.  In
Figure~\ref{fflxvpa} we plot the relative brightness, averaged
radially from 0.7 to $1.0 \, \thout$, as a function of p.a.\ for
several early epochs.  The p.a.\ of the gap clearly changes with time.
The most consistent evolution is for the period between $t = 264$~d
and 520~d, for which the gap moves from p.a.\ = 250\arcdeg\ to
375\arcdeg.  The peak is at p.a.\ = 100\arcdeg\ at $t = 264$~d and
moves to p.a.\ = 180\arcdeg\ by $t = 451$~d.  To further illustrate
this evolution, we plot in Figure~\ref{fminmaxpa} the p.a.s
of the peak and the gap in the ridge as determined from the images for
the period $t = 175$~d to 635~d.  This figure shows that the apparent
rotation is not uniform.  It is more pronounced for the gap, which
moves consistently counter-clockwise by $\sim 130$\arcdeg\ between $t
= 264$~d and 390~d. The peak rotates clockwise by $\sim 50$\arcdeg\ 
between $t = 390$ and 451~d, and tends to rotate slightly
counterclockwise at other times.  We discuss the
possible cause of this apparent rotation in \S\ref{sorigins} below,
but we note here that it is possible that the apparent rotation is
caused by a steady increase in brightness at p.a.\ of about
$-110$\arcdeg, and an unrelated decrease at p.a.\ $\sim 180$\arcdeg.
In other words the brightness at different p.a.'s may be evolving
independently, albeit with a timescale on the order of 200~d, and the
appearance of a rotation coincidental.

What is the significance of the apparent rotation?  The apparent
rotation is also suggested at 5~GHz where the first resolved image at
$t = 352$~d shows a very similar pattern, in particular having the
same orientation as that at 8.4~GHz.  From there on, the p.a.s of the
peak and the gap at 5~GHz track those at 8.4~GHz well (see
Figure~\ref{fsnmaps}).

More generally, we can assess the significance of changes apart from
homologous expansion between images of SN 1993J at different epochs by
scaling the images to the same effective size, and then comparing the
remaining differences between the images with the image uncertainty.
We find that the differences between the 8.4-GHz images at $t = 264$~d
and 451~d, when appropriately scaled in size and flux density and then
convolved to the same resolution, are six times the combined \sbg.
This apparent rotation thus represents a quite significant change in
structure even with the conservative assumption of an uncertainty of
1.4\sbg\ discussed in the previous section.

\subsection{Development of Three Hot Spots Along the Ridge from 686 Days 
to 1253 Days\label{stwospots}}

As the radio shell expands further, our relative resolution increases,
and more complex structure becomes visible. The opening in the ridge
narrows and the simple pattern with a single peak and gap gives way to
an almost closed shell structure with an evolving, complex modulation
along the ridge.  In particular, starting at $t = 774$~d, the
southeastern hot spot splits into an eastern and a southern one, and
gaps appear to the north and the southwest.  At $t = 1253$~d a large
arc has developed stretching over about 150\arcdeg\ in p.a., from the
northeast to the south.  During the same period, a third hot spot
develops in the east, but in contrast to the other two hot spots, this
hot spot remains stationary.

\subsection{The Filling of the Northern Opening in the Ridge from
1253 Days to 3345 Days}

From $t = 1253$~d on, the northern opening of the ridge
begins to fill in.  By $t = 1693$~d the original opening has
essentially vanished, 
and from $t = 2064$~d on the original opening has brightened to the
extent that it becomes a hot spot.  By $t = 2432$~d, there are again
three hot spots, a prominent one slightly east of north and ones in
the west-southwest and south-southeast.  At 8.4~GHz the dynamic range
has decreased to $\sim 10$ for the last two epochs, at which level the
brightness modulations along the ridge can no longer be discerned in
detail.  However, the pattern of modulation can be seen also in the
5.0 and 1.7~GHz images which have a higher dynamic range.  Our latest
image, at $t = 3345$~d and 5~GHz still exhibits the hot-spots to the
west-southwest and the south-southeast.  The hot-spot east of north
is also still present, but has broadened somewhat.

\begin{figure}
\epsscale{1.0}
\plotone{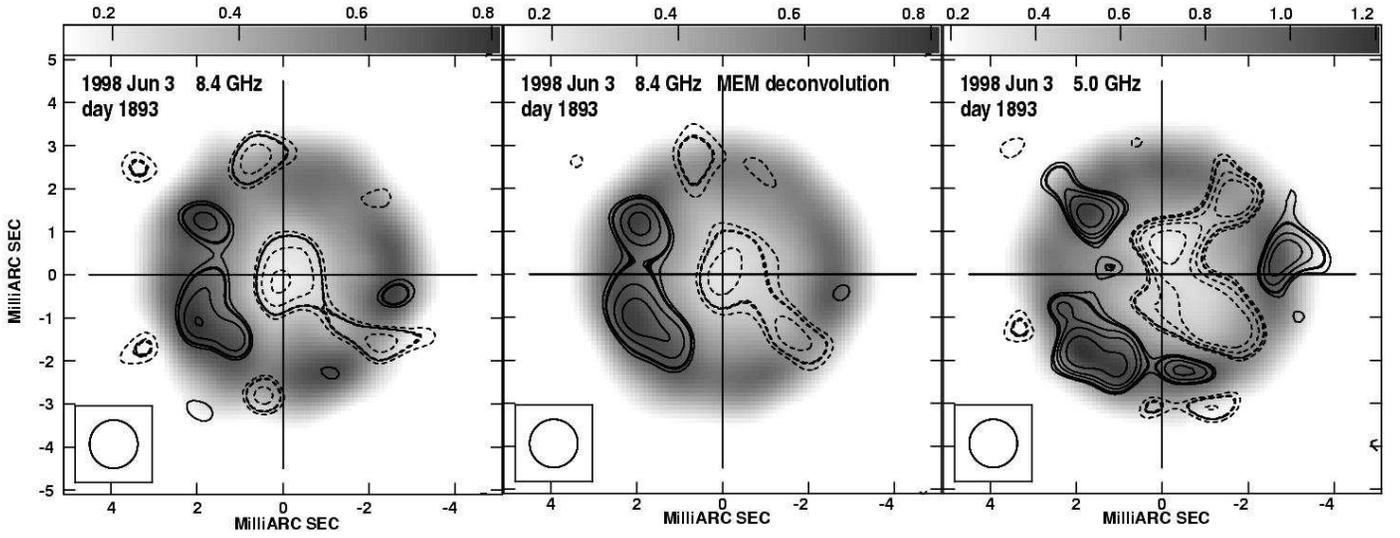}
\figcaption{The deviation of the brightness distribution of the radio
shell from the projection of a uniform spherical shell at $t =
1893$~d. The resolution is 1.12~mas.  The greyscales, labeled in
m\Jb, represent the brightness in the deconvolved image. The contours
represent, in units \sbg, the deviations from the best-fit spherical
shell model, also convolved to the same resolution.  They are drawn at
$-7$, $-5$, $-4$, {\bf $-$3}, $-2.5$, 2.5, {\bf 3}, 4, 5, and 7\sbg,
with the $\pm 3$\sbg\ contours being emphasized.  The true image
uncertainty is $1.2 \sbg \sim 1.4 \sbg$ (see \S~\ref{suncert}).  The
contours thus indicate the significance of deviations from a uniform
shell, with the extrema of both signs showing the most significant
deviations.  Features with amplitudes in excess of $\pm 3$\sbg\ are
likely to be real, while features with lower amplitudes could be, but
are not necessarily, due to image noise. The left panel shows the
8.4-GHz CLEAN image, the center panel shows an 8.4-GHz maximum entropy
image, biased to be as near to the uniform shell as allowed by the
data, and the right panel shows the 5-GHz CLEAN image.
\label{fsigmap}}
\end{figure}

Given the complexity of the structure, and the decreasing dynamic
range of the images, it may not necessarily be apparent which features
in the late-epoch images are real, and which are merely due to noise.
Since we found in Papers~I and II that the projection of a uniform
spherical shell, with a shell thickness of 20\% of the outer radius,
was a good overall description of the supernova, we will examine the
deviations of our images from such a model in order to illuminate the
significance of the features in the images.  As an example, in
Figure~\ref{fsigmap}, we show again our images for $t = 1893$~d at 8.4
and 5.0~GHz in grayscale, with contours showing the significance of
departures from a spherical shell.  More precisely, the contours
represent the deviations from the projected, convolved uniform
spherical shell model  of the image, in units of \sbg.

There are significant deviations from a uniform, spherical shell:
5\sbg\ and 9\sbg\ at 8.4 and 5.0~GHz respectively.  Note that while
the deviations at 5.0~GHz are more significant due to the higher
dynamic range at that frequency, they are not larger, that is the
supernova is not more uniform than at 8.4~GHz.  Note also that even
features with significance less than 3\sbg\ may well be real since the
addition of noise will not cause the disappearance of features at low
amplitudes. In this example, as in the other images, there is
excellent correspondence between the significant features at the two
frequencies, in particular in this case for the opening to the northeast and
the brightening to the southwest.

The CLEAN deconvolution process is subject to a known tendency to
introduce small scale corrugation into the images (Cornwell \& Braun
1988).  To test to what extent this tendency might be responsible for
the structure in our images, we made a biased image, again for $t =
1893$~d at 8.4~GHz, by deconvolving with maximum entropy (Cornwell
1988, 1999) using a default image consisting of the relevant projected
uniform spherical shell.  Even in this image, which has a relatively
low dynamic range, and is biased to be as close to the uniform shell
as is allowed by the data, there are deviations from a uniform shell
of up to 25\%, or 5\sbg, at 8.4~GHz.  Even at this relatively late
time, then, there are significant brightness deviations from a uniform
spherical shell.

\subsection{Polarization of SN~1993J \label{spoln}}

Synchrotron radiation is inherently polarized (e.g., Pacholczyk 1970),
and the polarization properties can potentially reveal the magnetic
field geometry.  The integrated polarization of SN~1993J is low, being
$< 1$\% at centimeter wavelengths, as determined from some of our VLA
observations (described Paper~II).  Because of the high
degree of circular symmetry, however, it is possible that the
polarization in a resolved image would be much higher.

In order to determine the polarization, observations in both senses of
circular polarization are required, which we obtained for all our VLBI
observing sessions after $t = 774$~d, with the exception of the one at
$t = 1107$~d (Paper II).  We did not detect any linear polarization
larger than the expected antenna polarization leakage of a few percent
in any of these VLBI sessions.

We carried out the full polarization calibration for several runs,
namely those of $t = 1356$~d at 8.4~GHz, of $t = 873$~d, 2064~d at
5~GHz, and of $t = 3164$~d at 1.7~GHz.  These runs were chosen because
the images had high dynamic range, and consequently are the best choices for
the search for polarization.  To correct for the polarization
leakage of each antenna, we determined the instrumental polarization
parameters from our observations of M81\*, which has very low
intrinsic linear polarization (Brunthaler et al.\ 2001) and is thus an
ideal instrumental polarization calibrator.

In no case did we find any significant polarization for SN~1993J\@.  We
give the
results in Table~\ref{tppol}.  In no case did we observe linear
polarization greater than the 3\%\footnote{The linearly polarized flux
density, $S_{\rm pol}$, is $\sqrt{Q^2 + U^2}$ where $Q$ and $U$ are the
measured Stokes parameters.  By definition, $S_{\rm pol}$ is positive,
and in the presence of noise is therefore biased.  This bias has been
calculated (Wardle \& Kronberg 1974), and a bias correction has been
implemented in the AIPS task COMB, which we used to calculate our
linearly polarized flux densities.  This procedure produces the
correct mean value of $S_{\rm pol}$, at the expense of occasionally
producing unphysical negative values.}
upper limit on the average polarization of the bright part of the ridge
is 4.4\% and on polarization at the image peak 9\%.

\section{A TIME-AVERAGED IMAGE OF SN~1993J\label{scompimg}}

Our last five 8.4-GHz images in Figure~\ref{fsnmaps} were convolved
with a CLEAN beam of FWHM 1.12~mas.  This is a somewhat lower
resolution than the $\sim 0.7$~mas obtained in the imaging process as
described in \S~\ref{sobsdat}.  However, their relatively low dynamic
range made it less useful to display them at the full resolution.  A
higher dynamic range, which would allow higher resolution, could be
achieved by properly averaging the images.  We showed in Paper~II that
the visibility curve of the radio shell, after scaling according to
the expansion, shows only small changes with time over the time
interval from $t = 996$~d to 2787~d. Since the radial profile of the
radio shell is the Hankel transform of the visibility curve, this
indicates, independent of any model-fitting and deconvolution, that
the average radial profile also shows only small changes with time.
Thus it is reasonable to average our data in time, accounting of
course for the overall expansion.  Such averaging allows us to study
those characteristics of the radio shell which don't change over the
averaging interval with higher angular resolution.

We accordingly averaged the data from the three latest epochs at
8.4~GHz to increase our dynamic range and form an image.  This image
will be an average over the three individual epochs, and we note that
in the presence of changes from epoch to epoch, some deconvolution
errors will be incurred.  Since, however, both the changes from image
to image and the sidelobes are only a fraction of the surface
brightness, such errors should be no more than a few percent of the
brightness.  Nonetheless, specific small-scale features in the
composite image should be interpreted with caution.

\begin{figure}
\plotone{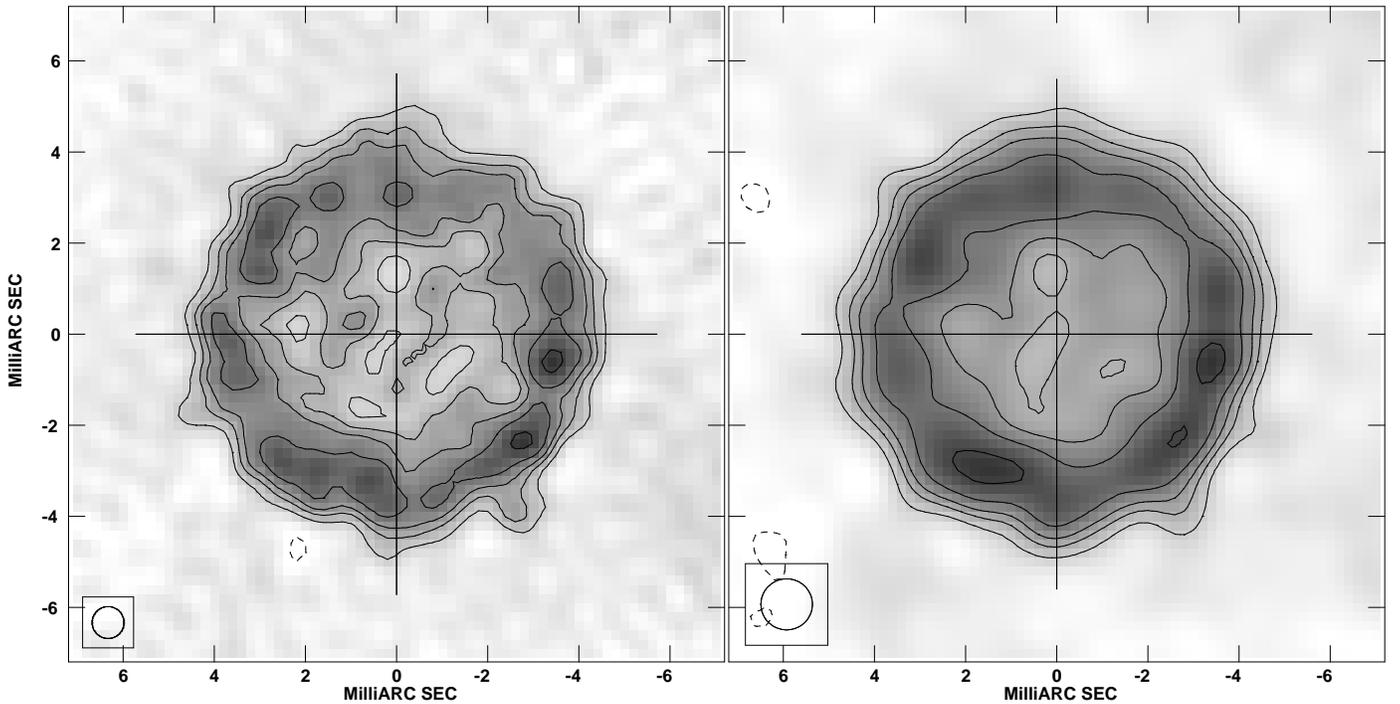}
\figcaption{Composite images made from the 8.4-GHz data at $t =
2080$~d, 2525~d and 2787~d (1998 December, 2000 February and
November), all aligned by the fit shell center, and scaled in flux
density and radius to the values of the $t = 2787$~d data (i.e.,
\thout = 4.49~mas; see text \S\ref{scompimg}, \ref{sshprof} for
details). The contours are drawn at $-16$, 16, 32, 45.3, 64 and 90\%
of the peak brightness, and \sbg\ was 5\% of the peak brightness on
both images.  On the left we plot an image with 0.70~mas resolution,
and on the right one with the 1.12~mas resolution used for the
individual epochs in Fig.~\ref{fsnmaps}.
\label{fcompimg}}
\end{figure}

To produce the composite image, we scaled and averaged the data from
the 8.4-GHz observing sessions at $t = 2080$~d, 2525~d, and 2787~d. To
account for the expansion of the radio shell, we normalized the
\uv~distances by scaling them by the outer radius of the fit shell,
normlized to the value at $t = 2787$~d (4.49~mas; see Paper~II).  For
example, if the supernova was $1.5\times$ smaller than at $t =
2787$~d, the \uv~distances would be divided by 1.5. We further scaled
the magnitudes of the complex correlation coefficients by the
respective total flux density for each epoch, and shifted their phases
so as to place the center of the fit shell at the phase center.  The
combined data set was then imaged and deconvolved as usual.  The
result is shown in Figure~\ref{fcompimg}.  The increased dynamic range
allows us to usefully image at the higher resolution of 0.7~mas.  The
shell appears very circular, but somewhat more uniform than it does in
the individual images (Fig.~\ref{fsnmaps}).  However, the brightness
still varies by a factor of $\sim 2$ between the brightest and the
faintest portions of the ridge.  These variations represent changes of
$\pm 4.5\sbg$, with the rms variation of the brightness around the
ridge being 2.2\sbg.  We estimated the errors in the deconvolution in
the same way we did for individual epochs in \S\ref{suncert} above,
and found that the uncertainty is somewhat higher in this case:
1.7\sbg.  The reason for the higher uncertainty is that the
\uv~coverage is less dense at the largest \uv~distances, which are
given higher weight in this image.  Despite the higher uncertainty, we
find that the variation of the brightness around the ridge is
significantly larger than the noise. In other words, there are likely
true brightness contrasts of $\gtrsim 1.4$ in the bright part of the
ridge which persist over periods $>1$ year.

We can also use this image to calculate the circularity of SN~1993J,
as we did for the individual images in Paper~I\@. The axis ratio of an
ellipse fit to the 20\% contour is 1.01.  The rms variation over all
p.a.'s of the radius of the 20\% contour is 4\%.  These values are
consistent with those we reported in Paper~I\@.
We note here that the 8.4~GHz image at $t = 2787$~d and the 5.0~GHz
images are circular to similar limits as reported for the earlier
8.4~GHz images in Paper~I\@.  On the latest image, at $t = 3345$~d
(5.0~GHz), an ellipse fit to the 20\% contour has an axis ratio of
1.06 (at p.a.\ = 76\arcdeg).  This is within the uncertainty for the
average ellipticity reported in Paper~I, but somewhat larger than
those obtained at other recent epochs.  Perhaps this is an indication
that SN~1993J is becoming elliptical.  Future measurements will be
required to confirm whether this is the case.

Does the radius variation in the outer contours reflect small-scale
departures from circularity, or is it due only to noise?  To determine
this we performed a Monte-Carlo simulation, substituting the Fourier
transform of a circularly symmetric model for the composite set of
visibilities above.  We then added a realistic amount of noise to
these model visibilities and imaged them.  Over numerous realizations
of the noise, we found that the average rms variation over all p.a.s
of the radius and the average ellipticity of the 20\% contour were not
significantly different than found for the real data.  The $1\sigma$
upper limit on the rms variation of the 20\% contour radius not
due to noise is 3\%.
We thus find no indication that the wavyness of the outer contours
represents any significant deviation from circularity.

There is, however, a small protrusion visible to the southwest on the
composite image.  At the 16\% contour, it extends about 0.6~mas beyond
the average radius.
It has a flux density of $\sim 0.5$\% of the total, equivalent to
$\sim 6 \sbg$.  There may be further, similar protrusions, but with
lower brightness, and therefore not distinguishable from the noise.
Is this protrusion visible in the individual images?  It is indeed
apparent in the 8.4~GHz images at $t = 2080$ and 2525~d.  At 5~GHz, it
is visible in the image at $t = 2432$~d, not in the one at $t =
2996$~d and only suggested in the last one at $t = 3445$~d.  It is not
apparent in the latest 1.7-GHz image at $t = 3164$~d
(Fig.~\ref{flimage}), but we might not expect it to be, given the
lower resolution at that frequency.  The protrusion is suggestive, but
its reality will have to be established by future observations.

\section{THE RADIAL BRIGHTNESS PROFILE\label{sshprof}}

\begin{figure}
\epsscale{0.75}
\plotone{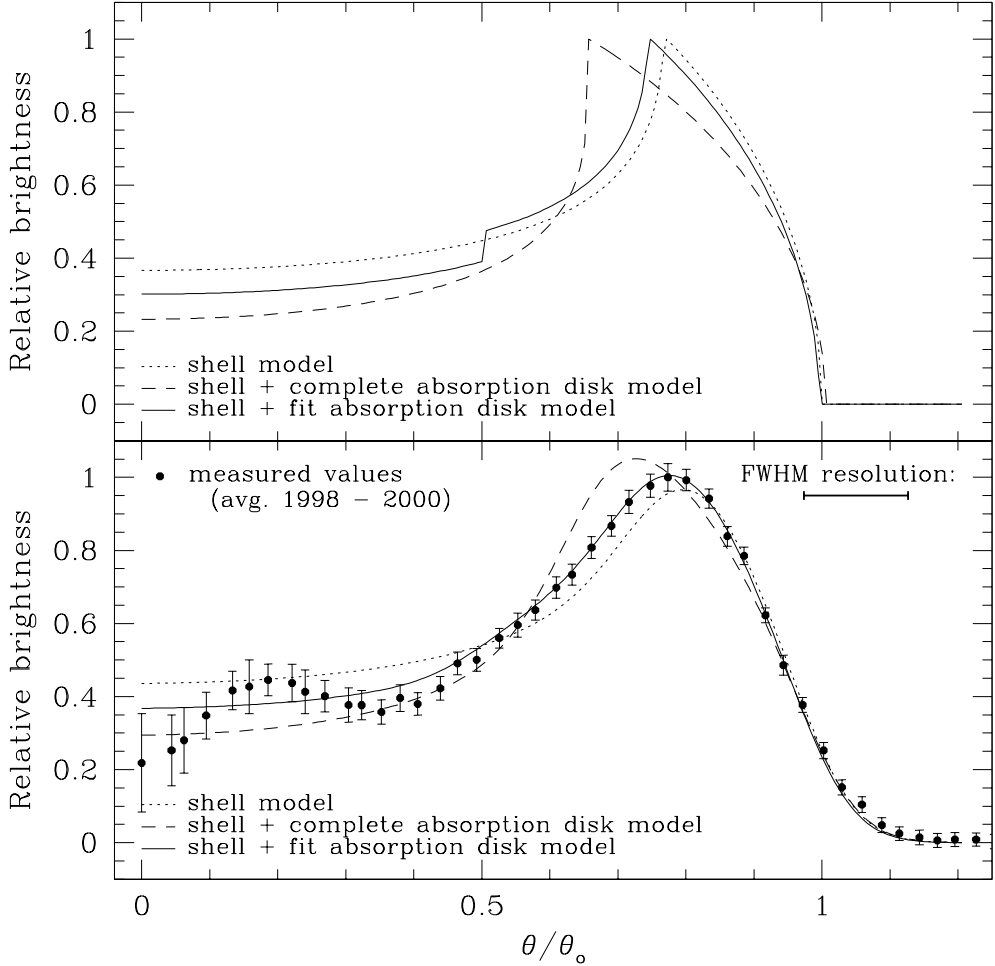}
\figcaption{Radial profile of the relative brightness, averaged over
all p.a.'s, versus angular radius, $\theta$, for the period 2080~d
$\leq t \leq 2787$~d.  We give $\theta$ in units of \thout, the outer
angular radius of the fit shell model.  The radial profile is that of
the composite image of Fig.~\ref{fcompimg}.  The profiles for several
models are also plotted.  The top panel shows the profile for three
unconvolved models.  The dotted line indicates the profile of the
fitted spherical shell model, the dashed line that of the fitted model
consisting of a uniform spherical shell with an absorption disk in the
center which represents a completely opaque interior of the shell, and
the solid line represents the best-fit model with a fitted absorption
disk, representing incomplete absorption in the interior of the shell
(\S\ref{sshprof}). The lower panel shows the measured average profile
and the profiles of the three models, all convolved to a resolution of
$0.16 \,\thout$. The uncertainties indicated are the standard errors of the
bin values, derived from the larger of the brightness uncertainty and
the standard deviation within the bin and accounting for the number
of beam areas within the bin.  The plotted values are correlated,
especially at small radii, because they are less than 1 beam width
apart.
\label{fcompprof}}
\end{figure}

It is of considerable interest to determine the average radial
brightness profile of the supernova.  The observed profile provides an
important constraint for modeling the shell structure, the density
profiles of the ejecta, and the emission and absorption processes.  To
study the profile of SN~1993J with the highest angular resolution we
used the composite image in Figure~\ref{fcompimg}, and produced a
profile of brightness {\em vs.}\/ radius, which we plot in
Figure~\ref{fcompprof}.  The uncertainties in the profile were derived
from the larger of \sbg\ and the rms scatter with position angle
within each radial bin, both divided by the square root of the number
of beam areas within each bin.

In addition, we plot the profiles of the models.  We used a spherical
shell of uniform volume emissivity as a model for our \uv~fits in
Papers I and II\@.  The radial profile of that model, fit to the present
composite data set is indicated by the dotted lines in
Figure~\ref{fcompprof}.  The fit toward the outer edge of the profile
is excellent, indicating that our model is good at least to the limit
imposed by the resolution.

In fact, the true radial profile of SN~1993J is not expected to
precisely follow this form (see e.g., Mioduszewski et~al.\ 2001; and
Jun \& Norman 1996a, 1996b).
We noted in Paper II (as did Mioduszewski et~al.\ 2001) that toward
the center of the shell, the observed profiles show systematic
deviations from that of a spherical shell of uniform volume
emissivity.  Specifically, Figure~\ref{fcompprof}\ shows that, from the
maximum inward, there is a deficit in emission toward the inside of
the ridge and then an excess in the center.  A thicker shell would
provide a better fit on the inside of the ridge, but a worse fit in
the center of the shell.  The deficit in the center can in general
also be seen in our well-resolved images at both 8.4 and 5~GHz, for
example those shown in Figure \ref{fsigmap}, where the deficit in the
center is still $> 4\sbg$ even in the biased maximum entropy image.

We will use the above scaled and time averaged data set from 2080~d
$\leq t \leq 2787$~d, which was used to make the time-averaged image
discussed in \S~\ref{scompimg}, to quantify these differences and to
better determine the real shell emission profile.  We will do this by
again directly fitting the \uv~data as described in Papers~I and II\@.
We note that by fitting directly to the \uv~data, we avoid
deconvolution errors so that our fit profile is a correct
representation of the time-averaged profile of the supernova.  Since
the most prominent difference between the model and our data seems to
be a deficit of emission in the center of the shell, we add a uniform
disk of {\em negative}\/ emission to our previous model of the
projection of an optically-thin, uniform spherical shell.  The
physical motivation for this parameterization, in particular for the
negative emission, is to represent possible absorption in the interior
of the shell, which might be expected on physical grounds (e.g.,
Mioduszewski et~al.\ 2001).  Such absorption would lower the
brightness in the central region of the projected supernova by
absorbing some of the flux from the rear of the shell.  Complete
absorption in the interior of the three-dimensional shell would imply
a reduction of exactly one-half the brightness in the central region,
corresponding to a negative emission disk whose brightness was the
negative of the observed average brightness in the central region.
The disk and the shell have the same center, but the radius and the
total (negative) flux density of the disk are free
parameters\footnote{We note that a disk is only an approximation to
the profile expected for absorption in the interior of a
three-dimensional shell.  The approximation is reasonable when, as in
our case, the radius of the absorbing material is well within the
outer radius of the shell.}.

We find that the best fit of the modified shell model gives a shell
thickness of 0.25 times the outer radius $\thout$ (equivalent to
$\thout / \thin = 1.34 \, \thout$), a radius for the absorption disk
of $0.5 \, \thout$, and an absorption of 4\% of the total flux
density.  The absorption corresponds to a reduction in the brightness
of $\sim 25$\% near the center of the shell.
The radial profile of this model is shown by the solid line in
Figure~\ref{fcompprof}.  The inclusion of an absorbing disk in the
center allows a slightly thicker shell to better fit the inner profile
of the ridge without an excess of emission in the center of the shell.
There are still some deviations of our measurements from the model
profile, especially near the center.  The resolution, however, is
0.70~mas (0.15 \thout), and therefore the several points nearest the
center which still show a deficit compared to the model are highly
correlated and probably not significant.

We repeated this analysis with the 5~GHz data sets of $t = 2992$ and
3345~d, and found a very similar result: the best fit of the modified
shell model gives a shell thickness of 0.25 times the outer radius
$\thout$ (equivalent $\thout / \thin = 1.33 \, \thout$), a radius for
the absorption disk of $0.4 \, \thout$, and an absorption of 4\% of
the total flux density.  At 5.0~GHz, the fit radius of the absorbing
disk is somewhat smaller than at 8.4~GHz.  For reasons we will
elaborate on below, however, the parameters of the absorbing disk
should be interpreted with caution.

In fact, at this early stage, the inner ejecta are still expected to
be quite opaque to radio waves (Chevalier 1982c; Reynolds \& Chevalier
1984; Mioduszewski et~al.\ 2001), and we might reasonably expect
almost complete absorption of the radio emission from the rear of the
shell.  Accordingly, we fit also a model with an absorbing disk
representing complete absorption and with a radius equal to the inner
radius of the shell.  In this case, the fitted shell thickness is
$0.35 \, \thout$. This model is also plotted in
Figure~\ref{fcompprof}, using a dashed line.  It clearly provides a
poorer fit than the model with the fitted absorption.

What is the uncertainty in the fit shell thickness?  The formal
uncertainty on the shell thickness when an incomplete absorption disk
is also fit is $0.03 \, \thout$.  This uncertainty is derived directly
from the visibility data, and therefore is not affected by the higher
on-source errors in the image plane discussed in \S~\ref{suncert}
above.
Since, however, our fit uses an approximation to the geometry of an
absorption disk, and since the geometry of the absorption probably
differs from a simple disk, our formal uncertainty may somewhat
underestimate the true uncertainty in the shell thickness\footnote{Our
uncertainty does take into account that the fitted shell thickness is
highly correlated with the fitted radius and total absorption of the
absorbing disk.  However, the presence of correlated visibility
errors, such as would arise from residual calibration errors, might
also slightly increase the true uncertainty.}.
In any case, the fits without any absorption disk and with a complete
absorption disk are clearly worse, suggesting that a very
conservative range for the true shell thickness is between $0.23 \,
\thout$ and $0.35 \, \thout$.

A more model-free estimate of the shell thickness can be derived from
the measured radial profile.  A three-dimensional shell emission
region, whose radial profile has sharp boundaries at \thin\ and
\thout\ like any of our three models in Figure~\ref{fcompprof} will,
when projected onto two dimensions, have a radial profile with
inflection points at \thin\ and \thout.  Convolution with the CLEAN
beam will smooth these inflection points. If we assume that extrema in
the second derivative of the convolved profile trace the location of
un-convolved inflection points, then these extrema occur at the
projected inner and outer radii of the shell.  We determined the
extrema of the second derivative from our measured profile in
Figure~\ref{fcompprof} numerically, and their locations suggest a
shell thickness close to the lower end of the above range, that is
$\sim 0.23 \, \thout$.  In summary, we think that our fit value of the
shell thickness of $(0.25 \pm 0.03) \thout$ is a reasonable estimate
of the thickness of the radio shell.

\section{DISCUSSION \label{sdicuss}}

With thirty-one epochs of observations of SN~1993J, phase-referenced
to the core of the host galaxy, we produced sequences of images of the
radio shell over more than nine years from the time of explosion to
the present.  In our Galaxy, radio shells of supernovae have been
observed over at most $\sim10$\% of their age.  SN~1993J has been
observed essentially over 100\% of its age.
In the first paper of this series, we determined the position of the
explosion center in the galactic reference frame with an accuracy of
about 160~AU, and placed an upper limit of 5.5\% on anisotropic
expansion in the plane of the sky.  In the second paper, we determined
the deceleration of the expanding supernova as a function of time,
together with the radio light curves and the spectrum, and interpreted our
results in terms of the interaction of the ejecta with the CSM\@.  In
this third paper we focus on the details of the images of the evolving
radio shell.

The earliest image shows an almost unresolved source with a radius of
520~AU.  The shell structure can be discerned from $t = 175$~d on.
The brightness along the ridge is modulated, with a maximum and a gap
located almost opposite.  This modulation pattern appears to rotate,
with the gap rotating from the west to the north-northeast in the
following 250~d.  From then on the structure along the ridge becomes
more complex.  Even at our last few epochs, however, the outer
contours of the (projected) radio shell remain circular within 4\%.
The brightness in the center of the shell is less than would be
expected of a uniform, optically thin shell.  With absorption in the
center taken into account, the ratio of the shell thickness is $0.25
\, \thout$.  No significant linear polarization was found.

These results are important for discussions about 1) the origins of
the brightness modulation of the shell and in particular, possible
structure in the ejecta, Rayleigh-Taylor instabilities, structure in
the CSM and a possible distant companion, and structure of the
magnetic field, 2) the relation between the outer edge of the radio
shell and the forward shock front, 3) a possibly non-uniform shell
thickness, 4) absorption in the center of the shell, and 5) the
possibility of discovering a pulsar nebula in the center of the radio
shell.  We discuss each of these aspects in turn.

We discuss our results in light of the theoretical understanding of
the structure of an expanding supernova shell described in the
introduction.  This structure is expected to comprise a forward shock,
driven outward into the CSM, and a reverse shock propagating inward in
the expanding ejecta (see e.g., Chevalier 1982a).  The CSM, in the
case of SN~1993J, consists of the slow, dense stellar wind of the
progenitor.  The contact discontinuity between the shocked CSM and the
expanding stellar envelope is located between the forward and reverse
shocks.  Radio emission is thought to be produced in the region
between the forward shock and the contact discontinuity by particles
accelerated by the shock and magnetic fields amplified by
instabilities at the contact discontinuity.  In the case of a
decelerating shell, it is expected that the contact discontinuity will
be Rayleigh-Taylor unstable (Gull 1973; Chevalier 1982a, b, c;
Chevalier \& Blondin, 1995), and that with time ``fingers'' of shocked
envelope material will extend into the shocked CSM\@.  The shear flow
along the sides of these fingers makes them subject also to the
Kelvin-Helmholtz instability, which may well further amplify the
magnetic field (Jun \& Norman 1996a, b).  We expect these
instabilities to apply to SN~1993J, which is already substantially
decelerated (Paper~II; see also Bartel et~al.\ 2000, Marcaide et~al.\
1997, 2002).  The magnetic field strength in SN~1993 can be estimated
from its size or velocity and the synchrotron luminosity and spectrum
(Paper~II; Chevalier 1998; Fransson \& Bj\"{o}rnsson 1998).
In fact, the magnetic field in SN~1993J is several orders of magnitude
higher than that expected in the stellar wind, which led Fransson \&
Bj\"{o}rnsson (1998) to argue that some form of field amplification
must be occurring as even a strong shock will amplify the field
only by a factor of four.

Many of the features of our images could be interpreted in a
straightforward way within this scenario: the forward shock is at or
near the outside boundary of the radio emission, and is highly
circular.  The contact discontinuity represents the inside boundary,
and shows deviations from circularity due to instabilities which are
visible chiefly as brightenings of the shell, rather than as actual
fingers, due to projection and to our limited resolution.  Numerical
modeling has shown that the Rayleigh-Taylor fingers do not generally
reach the forward shock (Chevalier, Blondin \& Emmering 1992; Jun \&
Norman 1996a, b), and so the outer edge of the shell remains highly
circular.

Our simple model of an optically thin, spherical shell of uniform
volume emissivity can serve as a rough description of a supernova with
this structure, with the inner and outer boundaries of the shell
being, respectively, the forward shock and the contact discontinuity.
Of course, the volume emissivity is not expected to be strictly
uniform within this region (see e.g., Mioduszewski et~al.\ 2001; Jun
\& Norman 1996a, b), but because even at the last epoch of our
observations at 8.4~GHz our resolution is only comparable to the shell
thickness, a uniform shell should provide an adequate first order
description.  However, our images clearly show deviations from this
structure.  We will discuss first the deviations of the azimuthally
averaged profile (Fig.~\ref{fcompprof}), and then the smaller-scale
modulations of the shell brightness visible in the images
(Figs.~\ref{fsnmaps}, \ref{flimage}, \ref{fcompimg}).

\subsection{Deficit of Emission in the Center of the Shell ---
An Opaque Interior?}

We found that our data show a deficit of emission in the center of the
shell, and an excess just inside of the ridge peak when compared to
the uniform spherical shell model with the best fit.  A thicker shell
would fit the ridge profile better, but produce an even brighter
central region.  We parameterized the deficit in the center by fitting
a disk of negative brightness in addition to the spherical shell
(\S\ref{sshprof}), and we found a deficit of $\sim 4$\% of the total
flux density, occurring over a radius of about half of the outer
radius of the shell, \thout.

There are a number of reasons why such a deficit might be expected for
SN~1993J, and below we discuss several of them, which are not
mutually exclusive.  An obvious one would be that SN~1993J be
intrinsically non-spherical, for example prolate, and that we observe
it face on.  However, since the outer contours are highly circular,
the prolate structure would have to be aligned with the line of sight
to within a few degrees. If we assume that the apparent deficit of
emission in the center is entirely caused by SN~1993J being in fact
prolate, the chance of an alignment close enough to the line of sight
so that SN~1993J would still appear as circular as it does is $< 5$\%.
Such a coincidence is unlikely.  Moreover, such an orientation is hard
to reconcile with the optical polarization data, which imply that we
are {\em not}\/ seeing SN~1993J along an axis of symmetry (Trammell
et~al.\ 1993; Tran et~al.\ 1997; see also H\"{o}flich et~al.\ 1996).

An equally obvious reason would be that the shell thickness is
different for different parts of the shell.  A shell which was
thinner, or had lower volume emissivity, at the front and/or rear than
it did along the circumference of the projected shell would produce
the observed deficit.  Once again, this would require a coincidental
alignment with the line of sight.  In view of the other plausible
explanations for the central deficit, we do not
consider these two possibilities further.

A third geometric explanation would be a shell which is in fact
thinner than the 25\% of its outer radius derived in \S\ref{sshprof},
but which is ``dimpled'', having local displacements from the
average radius.  In the center, a thinner shell seen in projection
would produce relatively lower brightness.  Near the limb, the
dimpling seen in projection would produce an apparently thicker shell
on average.  Such a geometry is not unlikely given the expected
Rayleigh-Taylor instability of the contact surface.

A fourth possibility is that the magnetic field is predominately
radial, as is expected if field amplification occurs due to
instabilities at the contact discontinuity (Jun \& Norman 1996a,
1996b), and as is observed in older remnants (Milne 1987; Dickel, van
Breugel, \& Strom 1991).  In this case even a spherical shell will
show enhanced emission along the limb because the magnetic field at
the limb will lie predominately in the plane of the sky, and the
synchrotron emission is strongest in directions perpendicular to the
magnetic field.  The polarization observations would appear to
contradict this scenario, since a well-ordered field should produce
high polarization which is not observed.  However, as we argue below,
internal Faraday rotation is likely to depolarize any centimeter
wavelength radio emission.

Finally, the most probable explanation is that significant absorption
in the interior of the shell is attenuating the emission from the side
away from us.  In fact, at this early stage, the inner ejecta are
still expected to be opaque to radio waves (Chevalier 1982c; Reynolds
\& Chevalier 1984; Mioduszewski et~al.\ 2001) because of their high
densities. Only if the inner ejecta have become transparent by
filamentation (Bandiera, Pacini, \& Salvati 1983) are we likely to be
able to see any emission from the distant side of the shell in the
first several decades after the explosion.

When we fit an absorbing disk to our data in addition to the spherical
shell, we found that the disk absorbed only 4\% of the total flux
density and had a radius of roughly half the outer radius
of the shell, \thout.  If the region inside the reverse shock were
completely opaque, one would expect an absorption of $\sim 25$\% of
the flux density, occurring over the inner radius of the shell or
$0.75 \, \thout$.
Even complete absorption over the fitted radius of the absorbing disk
would result in 9\% of the flux density being absorbed.  Thus, taken
at face value, our fit of the absorption disk suggests incomplete
absorption over a region well within the reverse shock, possibly
implying filamentation of the inner ejecta.

\begin{figure}
\epsscale{0.75}
\plotone{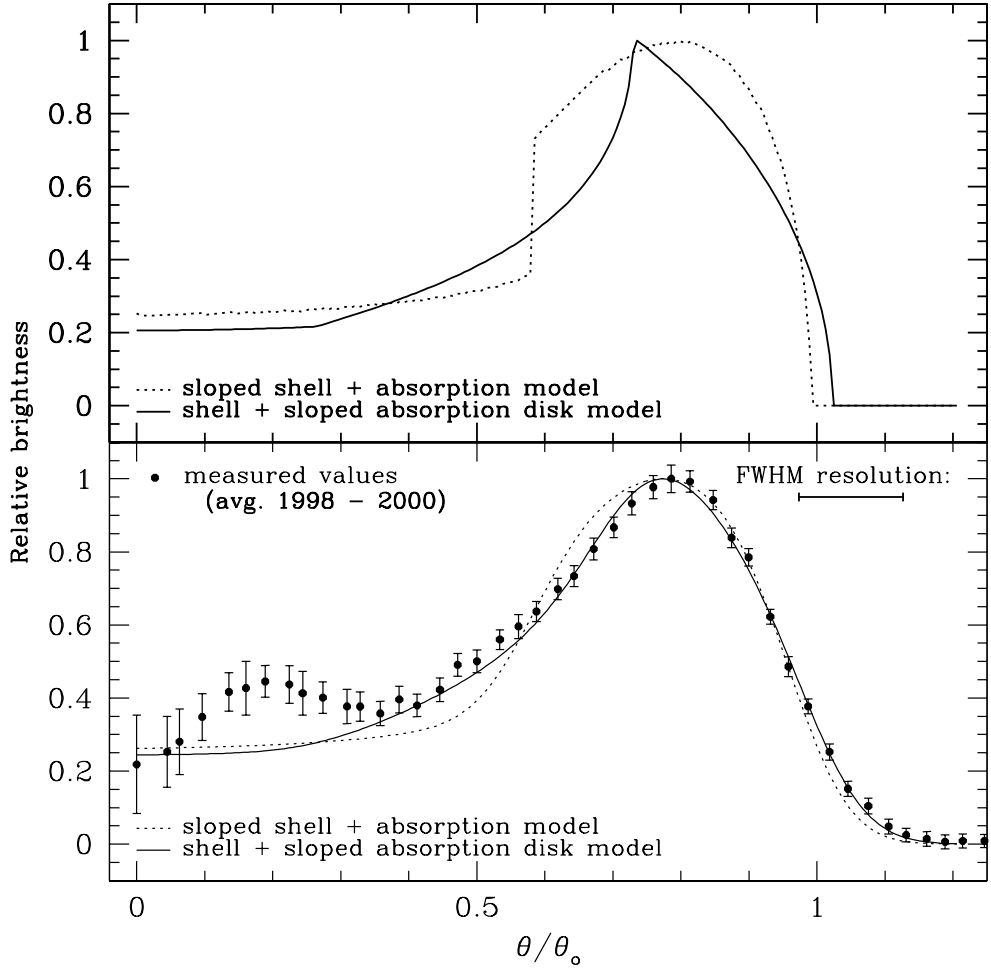}
\figcaption{As Figure~\ref{fcompprof} above, but two alternate models
are plotted.  The dotted line shows a model which has complete
absorption inside the inner radius of the shell, and volume emissivity
that is zero at the inner radius of the shell and then rises linearly
with radius till the outer radius of the shell. The inner radius was
$0.56\times$ the outer radius.  Such a model has the shallowest slope
possible inside the ridge line.  A good fit to the inside of the ridge
line cannot be obtained with such a model.
The solid line shows a model with and complete absorption in the
center, but a gradual transition to zero absorption at the inner
radius of the shell, but with uniform volume emissivity within the
shell.  Specifically, the fraction absorbed by the disk is 100\% from
the center up to a radius of $0.4\times$ the inner radius of the
shell, and decreases linearly with radius to 0\% at the inner radius.
The inner radius was $0.71\times$ the outer radius.  Note that this
fit is non-unique, hence the values of the model parameters are
consistent with but not demanded by the data.
\label{fslopepr}}
\end{figure}

Complete absorption at the inside radius of the shell would result in
a drop of 50\% of the surface brightness at the projected
inner radius or \thin.  Even with our limited resolution, this would
produce a steep drop in the profile to the inside of the ridge line,
as can be seen in the complete absorption model profile in
Figure~\ref{fcompprof} (dashed line).  The profile of that model is too
steep on the inside of the ridge-line.  In fact, even if we modify the
shell profile so that the volume emissivity of the shell is 0 at the
inner radius, and rises linearly to the outer radius, in other words,
to soften the slope inside of the ridge line as much as possible,
the model profile does not fit the observed one.  This profile is
plotted in Figure~\ref{fslopepr} (dotted line).

We conclude that if there is complete absorption in the interior of
the shell, then the outer edge of the absorbing region must be
somewhat soft, or that the optical depth near the inside edge is still
small, and it rises only gradually towards the center of the shell.
We plot such a model in Figure~\ref{fslopepr} (solid line). We note that
this model is not unique.  Rather than varying the optical depth, or
the opacity as a function of the radius, we vary the fraction
transmitted through the interior of the shell.  It seems likely that
this variation in effective opacity is due to mixing, being more
properly described as a variation in the filling factor of the opaque
material, in which case our treatment is reasonable.

However, the other mechanisms discussed above may also modify the
brightness.  Furthermore, a non-uniform volume emissivity in the shell
would also alter the apparent absorption derived by fitting a uniform
shell.  A more quantative determination of the emission and absorption
profile will have to wait for higher relative angular resolution and
the development of more sophisticated \uv~plane models.
In summary we believe some absorption in the interior of the shell is
strongly suggested by the data.  Complete absorption in the center of
the shell is compatible with our data, however, complete absorption
everywhere inside the inner shell radius is not.

\subsection{Does the Outer Edge of the Radio Shell Coincide with the Forward Shock?}

How is the outer edge of the radio shell related to the forward shock
front?  One would expect the radio emission region to be bounded on
the outside by the forward shock, which is principally responsible for
accelerating the electrons which produce the radio emission.  However,
since the magnetic field, also necessary for synchrotron emission, is
thought to be principally generated by amplification occurring near
the Rayleigh-Taylor unstable contact interface, it is possible that
the effective radio emission turn-on is significantly inside the
forward shock (e.g., Jun \& Norman 1996a).  Therefore the radio
emission itself can only indirectly reveal the location of the forward
shock.  Nonetheless, our radio images do give some suggestions
concerning both geometry and location of the forward shock.

First, Mioduszewski et al.'s (2001) hydrodynamical modeling of
SN~1993J gives a distance between the inner and outer shocks of $\sim
20$\% the radius of the outer shock\footnote{The self-similar
solutions of Chevalier (1982a) also predicts a shell thickness, and
give a thickness of 23\% for the values of $m = 0.83$ and an external
density $\propto r^{-2}$, typical for a few years after the explosion
(Paper II).  The predicted shell thickness varies only weakly for
different $m$ (Chevalier \& Fransson 1994). However, since we showed
in Paper~II that the evolution is not self-similar, it is possible
that the shell thickness may vary systematically from that determined
by Chevalier.}.  Our measured value of the radio shell thickness of
$25 \pm 3$\% of the outer radius is somewhat larger.
It seems unlikely that the radio emission originates either inside the
reverse shock or outside the forward shock.  It therefore seems most
plausible that the location of the forward and reverse shocks do
indeed coincide closely with \thout\ and \thin.

Second, the projection of the radio shell has remained remarkably circular
from the earliest observations to the present.  We determined an upper
limit to deviations from circularity of 5\% for the radio shell from
$t = 30$~d to 90~d (Bartel et~al.\ 1994) and of 3\% for later times
(Paper I; this paper), despite the significant modulations of the
brightness along the ridge.  The outer radius of the model radio shell
is very well determined and evolves very smoothly as a function of
time (Paper~II).  These characteristics suggest that, at least on
average, the outer edge of the radio shell is not influenced by local
peculiarities of the brightness distribution but determined by a
fundamental parameter of the expanding supernova.  It also suggests
that the forward shock, like the outer edge of the radio shell, is very
circular in projection, since it would be unlikely that any radio
``turn-on'' distance varies so as to make the radio outer edge
circular when the shock front is not.

Is the outer radio turn-on sharp?  Our model of a spherical shell with
uniform volume emissivity produces an excellent fit, especially near
the outer edge, to the projected brightness distribution (see
Figure~\ref{fcompprof}, Paper~II).  A distribution of volume
emissivity that drops sharply at the outer edge of the shell is
therefore consistent with our data.  However, a sharp outer edge to
the three-dimensional radio shell is not required by our data, since
our resolution is only comparable to the shell thickness.
Specifically, a model in which the volume emissivity is maximal at the
inner radius of the shell and then decreases linearly to zero at the
outer radius can be made consistent with the observed radial profile
from the maximum outwards. The limited resolution and projection onto
the sky render the recovery of the details of the distribution of
volume emissivity virtually impossible.  Future observations with a
higher relative resolution will be required to determine the details
of the variation of volume emissivity in the shell.

Comparison of the optical expansion velocities with the expansion
velocity determined from the radio observations could shed further light
on the relation between the outer edge of the radio emission and the
forward shock.  We will discuss this subject in more detail in a
forthcoming paper (Paper IV). 

\subsection{The Origins of the Brightness Modulation of the Shell
\label{sorigins}}

The brightness along the ridge of the projected shell is substantially
modulated at all epochs.  Even in our time-averaged image from 2080~d
$\leq t \leq 2787$~d, the brightness varies by a factor $\gtrsim 1.4$.
This variation in the brightness of the projected shell implies an even
larger variation in the volume emissivity of the radio shell.  There
are a number of possible causes for this variation.  We will discuss
in turn the four most likely causes and assess their relevance to the
structure apparent in our images and to its evolution in time.

\begin{trivlist}
\item{1. {\em Structure in the ejecta}}.  Any structure in the supernova
ejecta, for instance clumping or velocity anisotropy, could cause a
modulation of the brightness in the radio shell.  Recent numerical
studies (see M\"{u}ller, Fryxell, \& Arnett 1991; Fryxell 1994;
Burrows, Hayes, \& Fryxell 1995) suggest that the ejecta in type~II
supernovae develop strong hydrodynamic instabilities even before shock
breakout, just minutes after core collapse.  
The linear polarization found in the optical emission (Trammell
et~al.\ 1993), and perhaps also the asymmetry found in the optical
lines (Lewis et al.\ 1994; Spyromilio 1994), suggest asymmetry in the
ejecta for SN~1993J\@.  Aspherical models for the ejecta were
suggested to account for the optical polarization.  In particular,
oblate models with a major to minor axis ratio as large as 1.7 could
explain the polarization results.  The brightness distribution in our
early radio shell images is strikingly aspherical.  Perhaps this
asphericity was caused by structure in the ejecta.

Evidence for fragmented ejecta has been found in several Galactic and
extragalactic supernova remnants, including Cas~A (Braun, Gull, \&
Perley 1987; Anderson et~al.\ 1994), Tycho (Seward, Gorenstein, \&
Tucker 1983) and Kepler (Bandiera \& van den Bergh 1991), all of which
have remained relatively circular in overall appearance despite the
fragmentation of the ejecta.  Numerical simulations by Cid-Fernandes
et~al.\ (1996) show that the fragmentation of the ejecta is a
continuing process, with progressively smaller clumps forming in the
first few years as the supernova expands. There is some observational
evidence for the existence of clumping in the ejecta of SN~1993J
(Filippenko, Matheson, \& Barth 1994; Spyromilio 1994; Wang \& Hu
1994).

The $l = 1$ (see \S\ref{saspects}) modulation seen at early times,
however, cannot easily be explained by fragmentation, which is
expected to occur on scales small compared to the shell diameter.
Furthermore if anisotropy in the ejecta did occur on scales of the
shell diameter, we would expect a non-circular shell, which is
contrary to our results.  Such fragmentation in the ejecta may,
however, be the cause of the smaller-scale clumpiness seen in the
radio shell at late times.  The modulation of the shell brightness due
to the fragmentation in the ejecta would likely expand with the shell,
with a modulation pattern changing only slowly with time.  Such a
slowly changing structure is indeed observed in our sequence of images
after $t \sim 1500$~d.

\item{2. {\em Rayleigh-Taylor instabilities}}.  Instabilities in the
expanding shell will produce a modulated shell brightness.  As
mentioned above, the contact surface between the ejecta and the CSM is
expected to be Rayleigh-Taylor unstable.  Theory suggests that the
growth of the instability be such that small fingers appear first and
larger ones later (Gull 1973; Chevalier 1982c).  This suggestion is
supported by the numerical work of Jun \& Norman (1996a, b), who also
find that the scale of the Rayleigh-Taylor fingers is small in
comparison to the shell diameter.
As with fragmentation of the ejecta above, modulation of the shell
brightness on the scale of its radius is unlikely.  This instability
is also expected to grow with time.  Therefore this mechanism is
unlikely to explain the prominent $l = 1$ modulation seen at early
times.  By contrast, the growing amplitude and decreasing relative
scale size expected of the instabilities at the contact surface are a
good match for the modulation seen at late times.

Could the possible protrusion seen in the last images be a
Rayleigh-Taylor finger, protruding beyond the outer shock?  As we
mentioned earlier, the fingers are generally not expected to reach the
outer shock front (Chevalier, Blondin \& Emmering 1992; Jun \& Norman
1996a, b).  However, Jun, Jones, \& Norman (1996) showed that a clumpy
exterior medium enhances the growth of the Rayleigh-Taylor fingers,
and allows a few fingers to grow enough to penetrate the outer shock.
It is not clear, however, that such penetration could occur within the
first decade.  Future observations will show whether the protrusion
is, in fact, real, and whether it continues to grow.

\item{3. {\em Structure in the CSM and a possible distant companion}}.
The structure in the shell could also be due to pre-existing
structures in the CSM, which will generally influence the shell
brightness more directly than by enabling the growth of a few
Rayleigh-Taylor fingers beyond the shock front as discussed above.  (See
Jones et al.\ 1998 for a general discussion of the effect of the CSM
on an expanding supernova).  In fact, in SN~1993J, we have a unique
probe of the nature of the CSM of an evolved star.  Clumping
in the external medium has been invoked to explain the radio
light-curves of SN~1993J (Van Dyk et~al.\ 1994), although Fransson \&
Bj\"{o}rnsson (1998) have fit the radio light-curves without resort to
clumping.  Although the origins of possible clumps in the stellar wind
are not well known, it seems likely that the clumps would form at the
stellar surface with scales $\ll r_{\rm star}$, and that the clumps
would then tend to dissipate as the wind flows outward, since the
sound-speed in the wind is comparable to its expansion velocity.  Thus
one would expect the structure produced by a clumpy wind to scale in
size and diminish in intensity with distance from the progenitor, and
be characterized by $l > 1$ modulation.

Could structure in the CSM be responsible for the $l = 1$ modulation
seen early on, and in particular for the apparent rotation of the
modulation pattern between $t = 223$~d and 451~d (\S\ref{searly})?
Blondin, Lundqvist, \& Chevalier (1996) simulated the effect of a
supernova expanding into a CSM which had an axisymmetric density
gradient.  Such a density gradient might result from a density
distribution in the stellar wind which was axisymmetric about the
progenitor's rotation axis.  However, an axisymmetric pattern would
produce an $l = 2$ rather than an $l = 1$ modulation.

To produce the apparent rotation of the modulation pattern we see, a
roughly spiral pattern in the CSM would be required, with a size of
$\sim 3000$~AU\@.  Given the different rotation of the brightness peak
and gap, and the fact that the apparent rotation seems to have a
beginning and an end in time, a CSM geometry considerably more complex
than a simple spiral would be required to account for the brightness
modulation in detail.  A possibility, albeit somewhat speculative, is
that spiral structure in the CSM was produced by the colliding winds
of a binary.  The slow wind velocity of SN~1993J's progenitor of $\sim
10$~\kms, along with the pitch of the spiral pattern, implies a long
period for the binary of $\sim 6000$~yr.  This long period is not
compatible with the close binary which probably stripped much of
progenitor's hydrogen shell mass (Paper II), and therefore suggests a
distant tertiary component to the progenitor system.  We note that a
dust plume with spiral morphology of roughly this size was recently
seen near the red supergiant VY~CMa by Monnier et al.\ (1999), who
propose that it is caused by such a distant binary companion.

\item{4. {\em Structure in the magnetic field}}.
The magnetic field structure in the CSM can also influence the
synchrotron brightness of the shell, even in the absence of a
significant density inhomogeneity.  The magnetic field in the CSM is
not expected to be dynamically important, and therefore not likely to
deform the forward shock.  The synchrotron emissivity, however,
depends on the orientation of the magnetic field, being smallest when
the field lies parallel to the line-of-sight.  The orientation of the
CSM magnetic field may thus influence the appearance of our images.
Numerical studies (Jun \& Jones 1999; Jun \& Norman 1996b) have
confirmed that the magnetic field orientation is important in
determining the synchrotron brightness.  In fact, the simulations of
Jun \& Norman (1996b) showed that the orientation of the weak external
field, which is amplified by the forward shock, can have a significant
impact on the synchrotron emissivity.  A twisted field orientation,
for example, might account for the peculiar apparent rotation of the
brightness distribution seen between $t = 264$ and 451~d.  At
early times the field in the interaction region would still be
dominated by the shock-compressed pre-supernova field, while at later
times, it would become dominated by the component which has been
amplified and also randomized by the combined Rayleigh-Taylor and
Kelvin-Helmholtz instabilities.  The field orientation might well
contribute to both the early $l = 1$ and the late $l > 1$ modulation
of the shell brightness.

What can the observed linear polarization tell us about the structure
of the magnetic field?  We observed low values for the polarization of
the bright ridge of SN~1993J (\S\ref{spoln}), whereas the synchrotron
emission from a region of uniform magnetic field is expected to be
70\% polarized (e.g., Rohlfs \& Wilson 1996).
The low polarization of $<3$\% that we observed therefore suggests
some source of depolarization, or a highly disordered field.  Faraday
rotation along the line of sight within the emitting region can
cause depolarization of the radio emission even if the field is well
ordered.  Since the thermal electron densities and magnetic fields in
SN~1993J are high, we might expect significant internal Faraday
rotation.  For our purposes here, the rotation measure, $RM$, can be
taken as
$RM = 810 \, n_e \, B L$ rad~m$^{-2}$ where $n_e$ is the thermal
electron density in cm$^{-3}$, $B$ is the magnetic field in mG, and $L$
is the path length in pc.  In Paper~II we estimated the magnetic field
at $t = 3000$~d to be $B \sim 5$~mG (see also Fransson \&
Bj\"{o}rnsson 1998, who obtain a higher value).  The density can be
estimated from the mass within the interaction region of 0.3~\Msol\
(Paper II) and its volume calculated from our shell radii.  For $t =
3000$~d we obtain a value of $n_e \sim 10^4$~cm$^{-3}$.  
Armed with these estimates, we can calculate the path length which
would cause a $2\pi$ rotation of the polarization vector at 8.4~GHz to
be $10^{-4}$~pc.  This path length is the scale on which the radio
emission would become depolarized.  It also is very short compared to the
size of the supernova, corresponding to an angle of only $\sim
7$~\muas\ in our images.  Clearly, we expect the relatively large
magnetic fields and densities to highly depolarize any centimeter
wavelength radiation merely due to line-of-sight depolarization.
Although the $RM$ will decrease as the supernova expands, the
brightness and thus the signal-to-noise ratio will decrease also.
Unless there is a strong separation between the thermal and the
synchrotron-emitting relativistic particles, it is unlikely that
useful polarized radio emission will be seen in the near future.

\end{trivlist}

In summary, the brightness modulation in our images likely shows the
effects of both the fragmentation of the ejecta and the Rayleigh-Taylor
instability expected at the contact discontinuity.  However, there
also seems to be another source of strong modulation which is most
significant at early times, in particular modulation with scales
comparable to the radius of the shell, or $\sim 4000$~AU.  Given the
high degree of circularity seen later on, the strong modulation of
brightness at early times is unusual.  Neither fragmentation of the
ejecta nor the Rayleigh-Taylor instability is likely to produce such a
modulation of the shell brightness.  The most probable candidates are
either density structures or an ordered magnetic field in the CSM.

\subsection{A Pulsar Nebula?}

SN~1993J is thought to have produced a compact remnant.  Given the
estimates of the progenitor mass, this compact remnant is probably a
neutron star rather than a black hole, and the neutron star is
generally expected to manifest itself as a pulsar.  The strong
relativistic wind of a young pulsar is expected to produce a bright,
flat-spectrum synchrotron nebula.  The expansion rate of such a pulsar
nebula is expected to be $<10$\% that of the shell.  Therefore even at
our latest epoch the angular size of such a nebula would likely be
$<1$~mas (Reynolds \& Chevalier 1984; Chevalier \& Fransson 1992; Jun
1998).  We find no such compact feature in our individual images (see
Fig.~\ref{fsnmaps}).  In particular, we examined a high-resolution
image at $t = 2787$~d, which had a convolving beam area of
0.48~mas$^2$, and found no emission near the center brighter than
0.11~m\Jb.
We also do not find any such compact feature in our composite image
from 2080~d $\leq t \leq 2787$~d, where there is a deficit in the very
center of the image, and the brightest feature in the central region
represents only about 0.05~m\Jb\ for a beam area of 0.55~mas$^2$.
At 8.4~GHz and the distance of M81, the equivalent brightness of the
$\sim 950$-year-old Crab Nebula would be 0.15~m\Jb.  A seven-year-old
pulsar nebula, on the other hand, is expected to be considerably
brighter than one 950 years old (Bandiera, Pacini, \& Salvati 1984),
and should therefore be detectable in our images even if it were
considerably less luminous than the Crab.  However, as mentioned
above, the material in the center of the radio shell is likely to
still have a high enough density so as to be opaque to radio waves
(Chevalier 1982c; Reynolds \& Chevalier 1984; Mioduszewski et~al.\
2001).  The non-detection of the expected pulsar nebula therefore
leads us to conclude that either the pulsar nebula is much weaker than the
Crab, or that the material immediately surrounding the putative pulsar
nebula cannot yet have become transparent by filamentation.

\section{CONCLUSIONS}

\begin{trivlist}

\item{1.} A sequence of VLBI images of SN 1993J at 30 epochs from $t =
50$~d to 3345~d shows the dynamic evolution of the expanding radio
shell of an exploded star in detail.

\item{2.}  The shell structure first becomes visible at $t = 175$~d,
and is present in all subsequent images.

\item{3.} The brightness distribution changes significantly and
systematically throughout the sequence of images.  The evolution is
clearly not self-similar.

\item{4.}  At $t = 175$~d, the brightness around the ridge is
significantly modulated, with a maximum or peak to the southeast and a
minimum or gap to the west.  Over the next 350~d, this pattern rotates
counter-clockwise, although the peak and gap appear to rotate by
slightly different angles and at slightly different times.

\item{5.} From $t = 774$~d to 1253~d, the gap fills in and three
hot spots develop to the west, south and east.

\item{6.} From then on the northern part changes significantly.  From
$t = 2080$~d, hot spot develops in the north-northeast at the p.a.\ of
the previous gap.  There are also two further hot spots, one to the
south-southwest and one to the west.

\item{7.}  Throughout our observing interval, the brightness around
the ridge of the projected shell is modulated by a factor of $\gtrsim
1.4$ on spatial scales of $\sim 4000$~AU (1~mas).

\item{8.} At early times, the modulation of the shell brightness on
the scale of the shell radius is substantial, whereas at later times,
the modulation on the scale of the radius is small.

\item{9.} Despite the modulation in brightness, even at our last
epoch, we find no evidence that the outer contours of the projected
shell deviate from circularity by more than a few percent.  However, a
first indication of a possible protrusion was found in our last
images.

\item{10.} The radio emission of the bright part of the shell is
$<4.4$\% linearly polarized.  Internal Faraday rotation is expected to
produce essentially complete depolarization.

\item{11.} The brightness in the central region of the projected radio
shell is lower than that of a model with uniform volume emissivity in
an optically thin, spherical shell.  We consider absorption in the
interior of the shell more likely than a non-spherical geometry with a
thinner shell at the front and/or the back. 

\item{12.} The outer edge of the projected shell accurately matches
that of the above model, thus an emission volume with a sharp outer
edge is consistent with our data.

\item{13.} The best fit value of the shell thickness, derived from
model-fits to the 8.4~GHz \uv~data between $t = 2080$~d and 2787~d and
by allowing for absorption in the center, was $25 \pm 3$\% of its
outer radius.

\item{14.} We detect no compact feature, which might be identified as
a pulsar nebula, at or near the center of the shell at any epoch.  At
8.4~GHz, we can place a limit of 0.11~mJy on the flux density of any
such feature at $t = 2787$~d, and a limit of 0.05~mJy on average for
the interval 2080~d $\leq t \leq 2787$~d.  These limits correspond to
0.7 and 0.3 times the current spectral luminosity of the Crab Nebula.
We conclude either that any pulsar nebula in the center of SN~1993 is
considerably less luminous than the Crab or that the interior of the
shell cannot yet have become transparent by filamentation.

\end{trivlist}

\acknowledgements

ACKNOWLEDGMENTS.  We thank V. I. Altunin, A. J. Beasley, W. H. Cannon,
J. E. Conway, D. A. Graham, D. L. Jones, A. Rius, G. Umana, and
T. Venturi for help with several aspects of the project.  J. Cadieux,
M. Keleman, and B. Sorathia helped with some aspects of the VLBI data
reduction during their tenure as students at York. We thank NRAO, the
European VLBI Network, and the NASA/JPL Deep Space Network (DSN) for
providing exceptional support for this extended and ongoing observing
campaign. We also thank Natural Resources Canada for helping with the
observations at the Algonquin Radio Observatory during the first years
of the program.  Research at York University was partly supported by
NSERC\@.  NRAO is operated under license by Associated Universities,
Inc., under cooperative agreement with NSF\@.  The European VLBI
Network is a joint facility of European and Chinese radio astronomy
institutes funded by their national research councils.  The NASA/JPL
DSN is operated by JPL/Caltech, under contract with NASA\@.  We have
made use of NASA's Astrophysics Data System Abstract Service.

\clearpage

\newcommand{\srr}{\tablenotemark{d}}
\newcommand{\sro}{\tablenotemark{e}}
\begin{deluxetable}{r@{ }l@{ }l r c c c@{\hspace{0.6in}} c c c}
\tablecaption{Parameters of the SN~1993J Images at 8.4 and 5 GHz\label{tsnmaps}}
\tabletypesize{\footnotesize}
\tablehead{ 
\multicolumn{3}{c}{Date} & \colhead{Age\tablenotemark{a}}  &
   \multicolumn{3}{c}{8.4 GHz} & \multicolumn{3}{c}{5.0 GHz} \\
 & & & & \colhead{Resolution\tablenotemark{b}} & \colhead{Peak} & \colhead{\sbg\tablenotemark{c}~}
       & \colhead{Resolution\tablenotemark{b}} & \colhead{Peak} & \colhead{\sbg\tablenotemark{c}~} \\
 & & &                  &        & \colhead{brightness} &  
                        &        & \colhead{brightness} \\
& & & \colhead{~(days)} & \colhead{(mas)}      & \colhead{(m\Jb)}     & \colhead{(m\Jb)} 
                        & \colhead{(mas)}      & \colhead{(m\Jb)}     & \colhead{(m\Jb)}
}
\startdata
 1993&May&17 &  50 & 0.45\srr & 45.5\phn & 0.30\phn  \\
 1993&Jun&27 &  91 & 0.46\srr & 83.3\phn & 0.36\phn  
                   & 0.80\srr & 61.7\phn & 1.28\phn \\
 1993&Sep&19 & 175 & 0.47\sro & 23.3\phn & 0.27\phn  \\ 
 1993&Nov& 6 & 223 & 0.49     & 15.6\phn & 0.11\phn \\
 1993&Dec&17 & 264 & 0.50     & 12.1\phn & 0.10\phn \\
 1994&Jan&28 & 306 & 0.54     &\phn 9.83 & 0.24\phn \\
 1994&Mar&15 & 352 & 0.57     &\phn 6.23 & 0.15\phn 
                   & 0.85     & 19.80    & 0.17\phn \\
 1994&Apr&22 & 390 & 0.60     &\phn 6.25 & 0.13\phn
                   & 0.88\sro & 18.75    & 0.30\phn \\
 1994&Jun&22 & 451 & 0.69     &\phn 6.32 & 0.073 
                   & 0.90\srr & 13.57    & 0.21\phn \\
 1994&Aug&30 & 520 & 0.73     &\phn 4.81 & 0.14\phn 
                   & 0.95     & 10.84    & 0.10\phn \\
 1994&Oct&31 & 582 & 0.74     &\phn 4.73 & 0.14\phn 
                   & 1.00\srr & 10.45    & 0.23\phn \\
 1994&Dec&23 & 635 & 0.78     &\phn 4.17 & 0.10\phn
                   & 1.01     &\phn 8.79 & 0.14\phn \\
 1995&Feb&12 & 686 & 0.80     &\phn 3.66 & 0.061 \\
 1995&May&11 & 774 & 1.00     &\phn 4.66 & 0.25\phn \\
 1995&Aug&18 & 873 &          &          &          
                   & 1.03     &\phn 3.87 & 0.084 \\
 1995&Dec&19 & 996 & 0.93     &\phn 1.90 & 0.088    
                   & 1.04\srr &\phn 3.86 & 0.20\phn \\
 1996&Apr& 8 &1107 & 0.97     &\phn 1.60 & 0.087
                   & 1.05\sro &\phn 2.91 & 0.11\phn \\
 1996&Sep& 1 &1253 & 1.02     &\phn 1.62 & 0.065
                   & 1.05     &\phn 2.28 & 0.040 \\
 1996&Dec&13 &1356 & 1.05     &\phn 2.59 & 0.056 
                   & 1.05     &\phn 1.81 & 0.034 \\
 1997&Jun& 7 &1532 & 1.10     &\phn 1.20 & 0.064 \\
 1997&Nov&15 &1693 & 1.12     &\phn 1.00 & 0.042
                   & 1.12     &\phn 1.12 & 0.034 \\
 1998&Jun& 3 &1893\tablenotemark{f} & 1.12     &\phn 0.91 & 0.029 
                   & 1.12     &\phn 1.20 & 0.031 \\
 1998&Nov&20 &2064 &          &          & 
                   & 1.12     &\phn 0.97 & 0.021 \\
 1998&Dec& 7 &2080 & 1.12     &\phn 0.88 & 0.038 \\
 1999&Jun&16 &2271 &          &          & 
                   & 1.12     &\phn 0.84 & 0.040 \\
 1999&Nov&23 &2432 &          &          &      
                   & 1.12     &\phn 0.71 & 0.037 \\
 2000&Feb&25 &2525 & 1.12     &\phn 0.48 & 0.040 \\
 2000&Nov&13 &2787 & 1.12     &\phn 0.32 & 0.030 \\
 2001&Jun&10 &2996 &          &          &
                   & 1.12     &\phn 0.42 & 0.016 \\
 2002&May&25 &3345 &          &          &      
                   & 1.12     &\phn 0.34 & 0.019 \\
\enddata
\tablenotetext{a}{Time in days since shock breakout on 28.0 March 1993.}
\tablenotetext{b}{Effective angular resolution, i.e., FWHM of the circular Gaussian convolving beam.}
\tablenotetext{c}{Standard deviation of the background brightness, see text 
\S\ref{suncert}.}
\tablenotetext{d}{Super-resolved, by less than 50\%.}
\tablenotetext{e}{Super-resolved along the long axis of the beam only, CLEAN beam preserves area of the dirty beam.}
\tablenotetext{f}{The 5.0 GHz observations actually occurred 1 day earlier.}

\end{deluxetable}

\begin{deluxetable}{l r c c @{\hspace{40pt}} c c @{\hspace{40pt}} c c}
\tablecaption{Percentage Linear Polarization of SN~1993J\label{tppol}}
\tabletypesize{\footnotesize}
\tablehead{
\colhead{Date} & \colhead{$t$\tablenotemark{a}}
 & Frequency & Resolution\tablenotemark{b}
 & \multicolumn{2}{c}{Mean Polarization for~~~} 
 & \multicolumn{2}{c}{Polarization of~}  \\
 & & &
 & \multicolumn{2}{c}{$I > 60$\% of peak \tablenotemark{c,d}~~~} 
 & \multicolumn{2}{c}{Image Peak\tablenotemark{d}~} \\
 & & & & value & $3\sigma$ upper limit & value & $3\sigma$ upper limit \\
 & (d) & (GHz) & (mas) & (\%)  & (\%)  & (\%)  & (\%)
}
\startdata
1995 Aug 17 &  873 & 5.0 & 0.93 & $0.2 \pm 0.7$ & 2.3 & $+0.5 \pm 1.7$ & 6 \\
1996 Dec 13 & 1356 & 8.4 & 1.29 & $1.1 \pm 1.1$ & 4.4 & $+2.9 \pm 2.1$ & 9 \\
1998 Nov 20 & 2064 & 5.0 & 0.96 & $1.2 \pm 0.9$ & 3.9 & $-2.1 \pm 2.4$ & 5 \\
2001 Nov 26 & 3164 & 1.7 & 2.61 & $0.5 \pm 0.7$ & 2.6 & $-0.8 \pm 1.7$ & 4 \\
\enddata
\tablenotetext{a}{The number of days since shock breakout on 28 March 1993.}
\tablenotetext{b}{The angular resolution (geometric mean of the FWHM
of the major and minor axes of the Gaussian restoring beam).  The
weighting used in imaging was chosen to maximize the signal-to-noise
ratio, therefore the beam size may differ from the corresponding one
in Table~\ref{tsnmaps} and Figure~\ref{fsnmaps}.}
\tablenotetext{c}{The (scalar) mean percentage linear polarization
over the region where the brightness in Stokes $I$ is $>60\%$ of the
image peak.}
\tablenotetext{d}{All percentage polarizations are corrected for the
noise bias.  They are therefore correct on average, but sometimes
unphysically negative; see text \S\ref{spoln}.}
\end{deluxetable}

\vfill\eject

\begin{figure}
\plotone{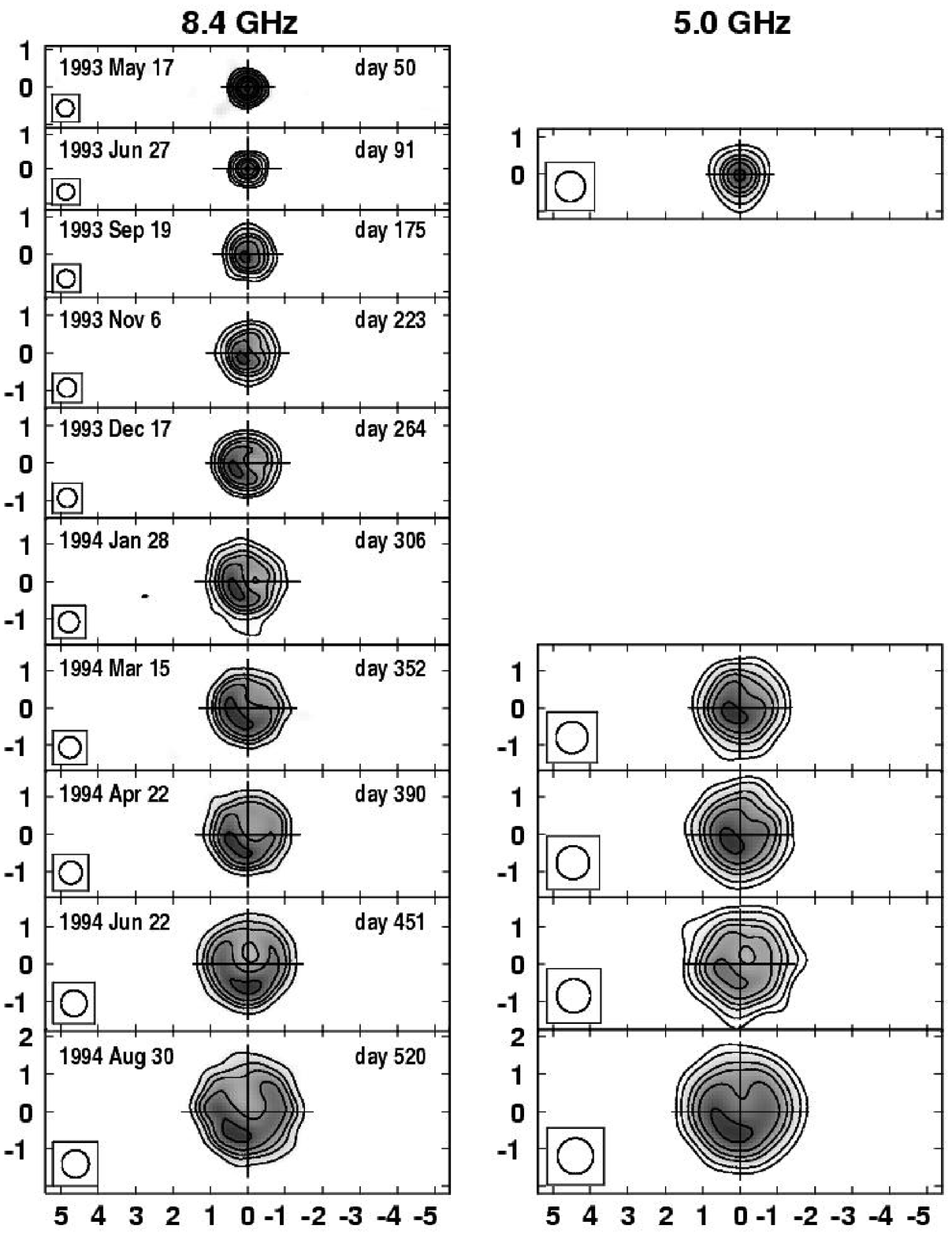}
\end{figure}

\begin{figure}
\plotone{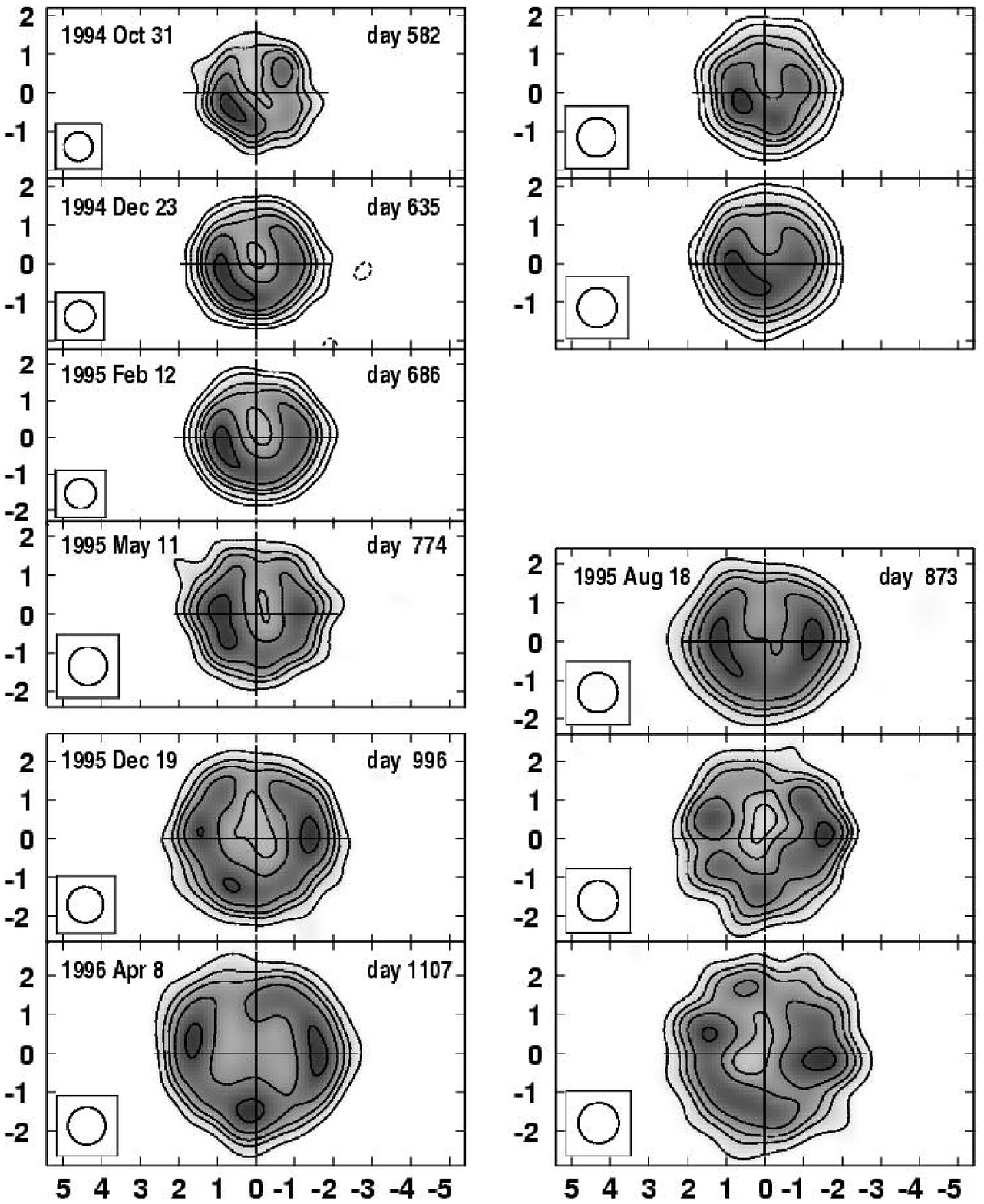}
\end{figure}

\begin{figure}
\plotone{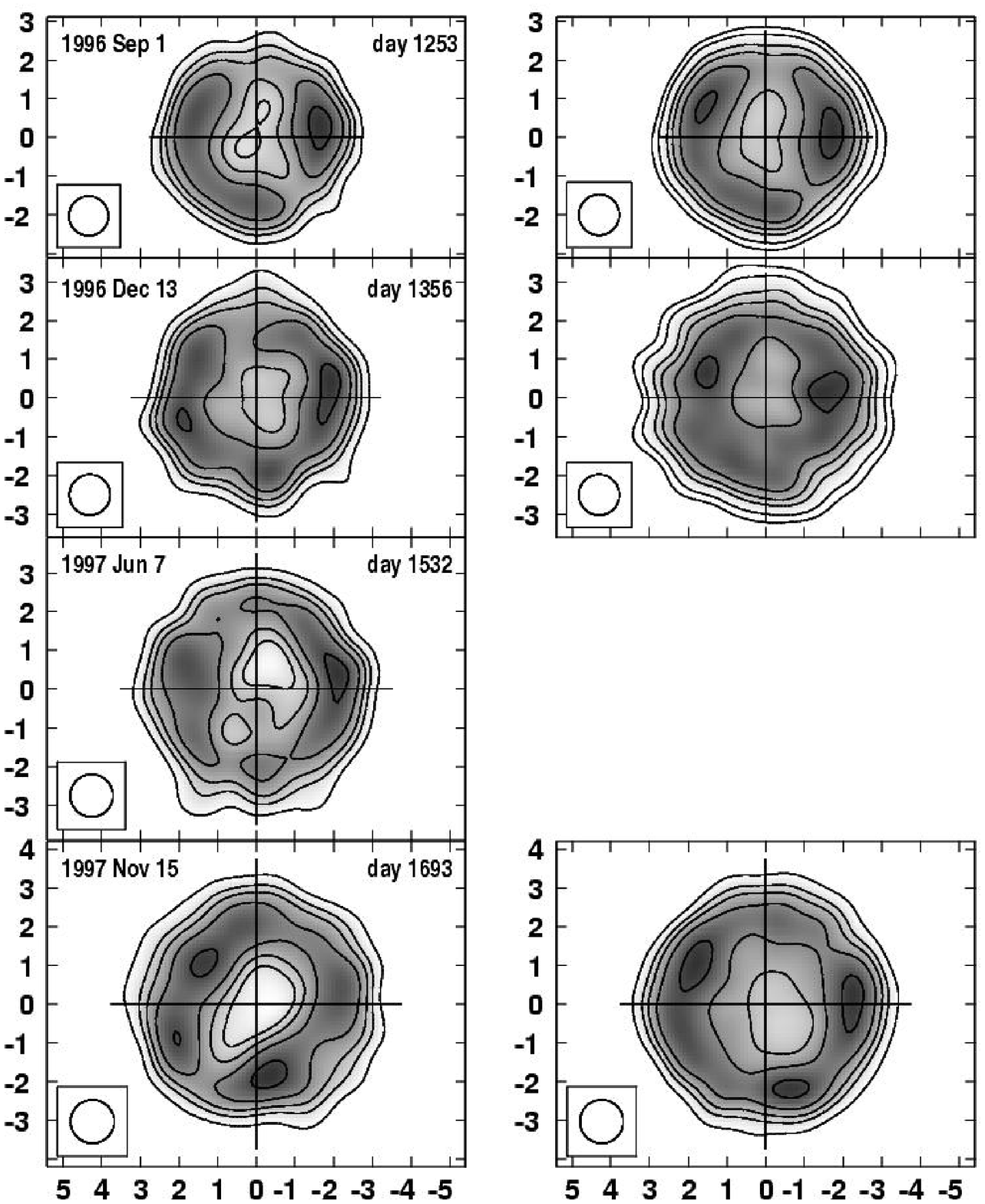}
\end{figure}

\begin{figure}
\plotone{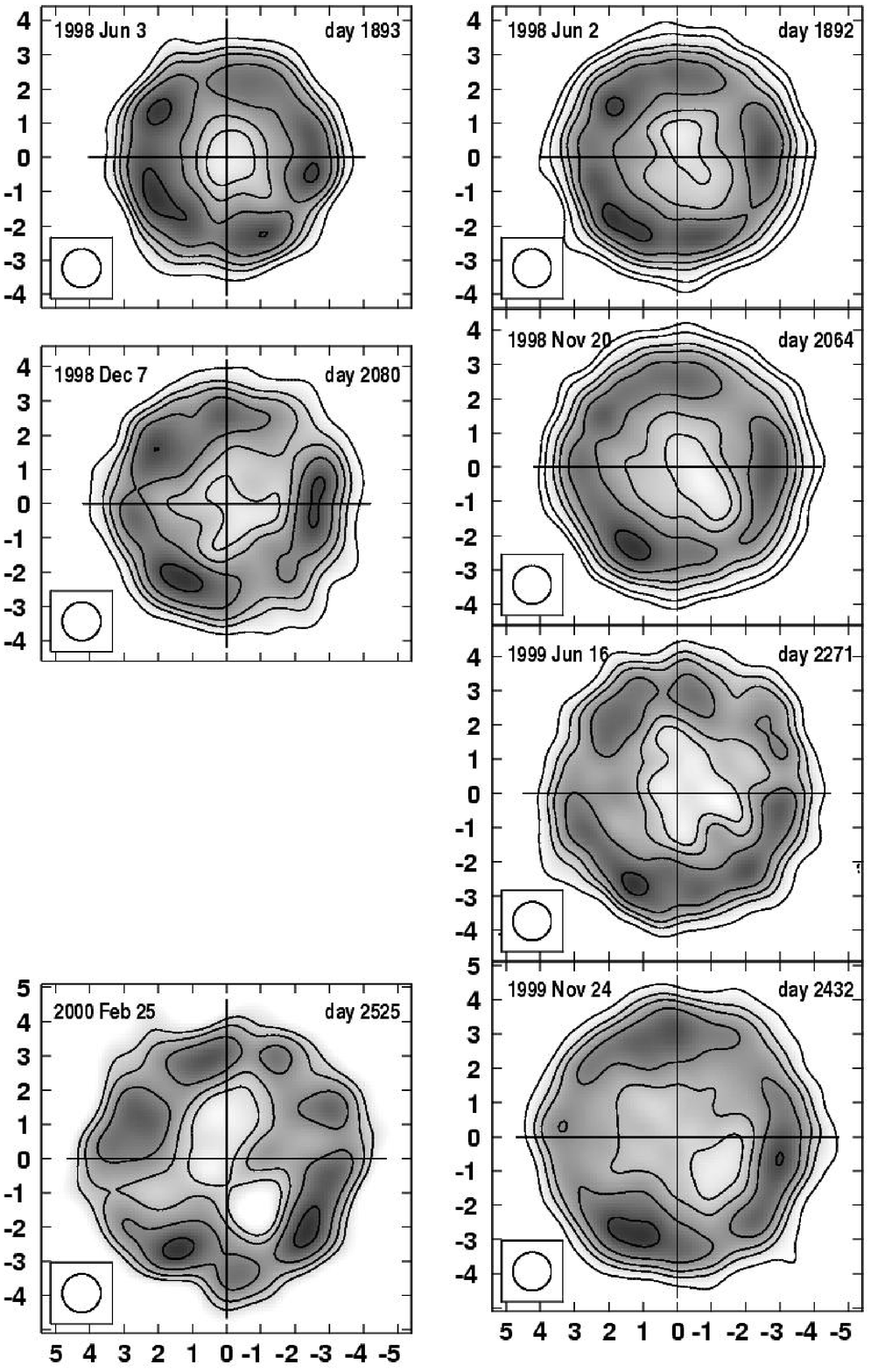}
\end{figure}

\begin{figure}
\plotone{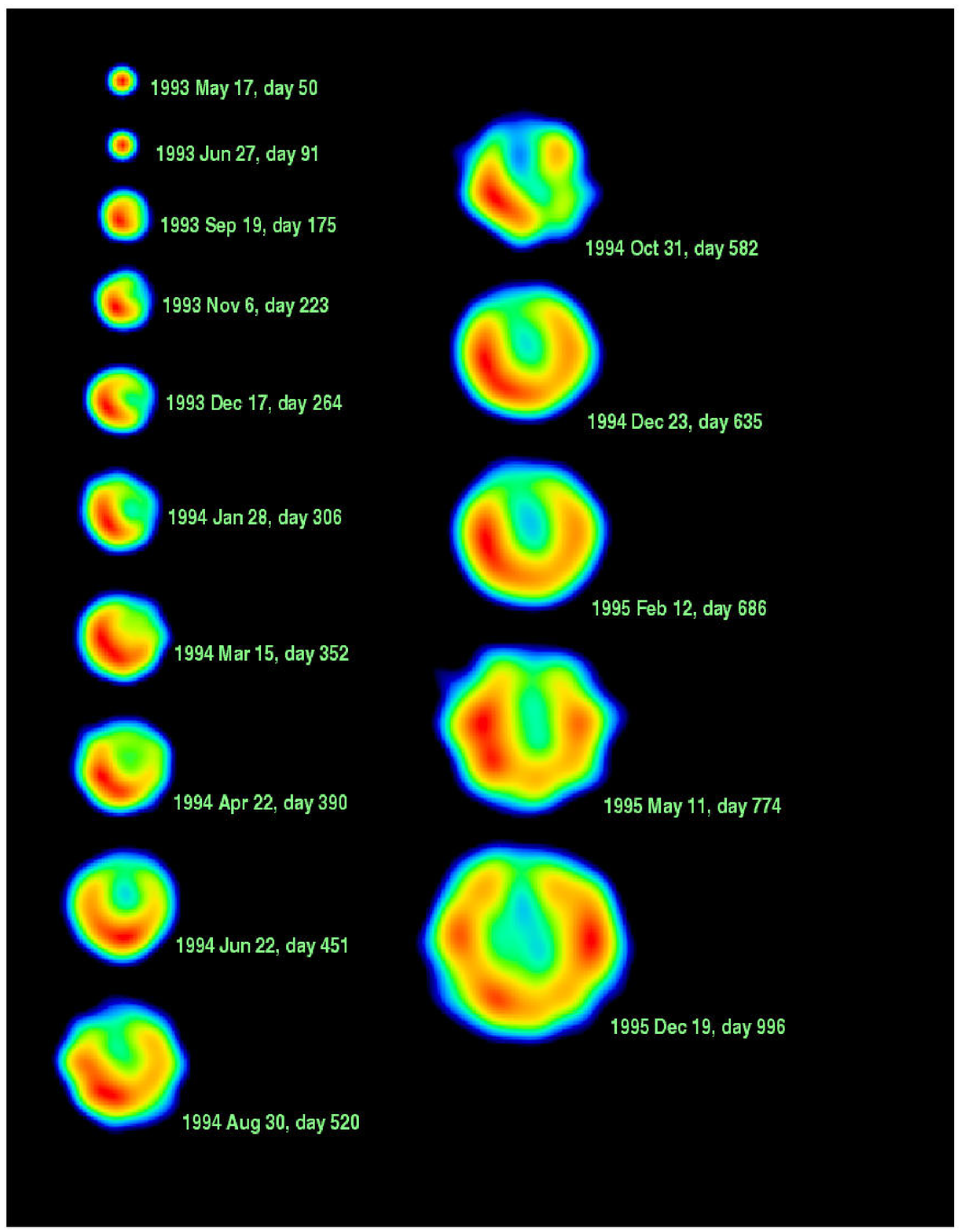}
\end{figure}

\begin{figure}
\plotone{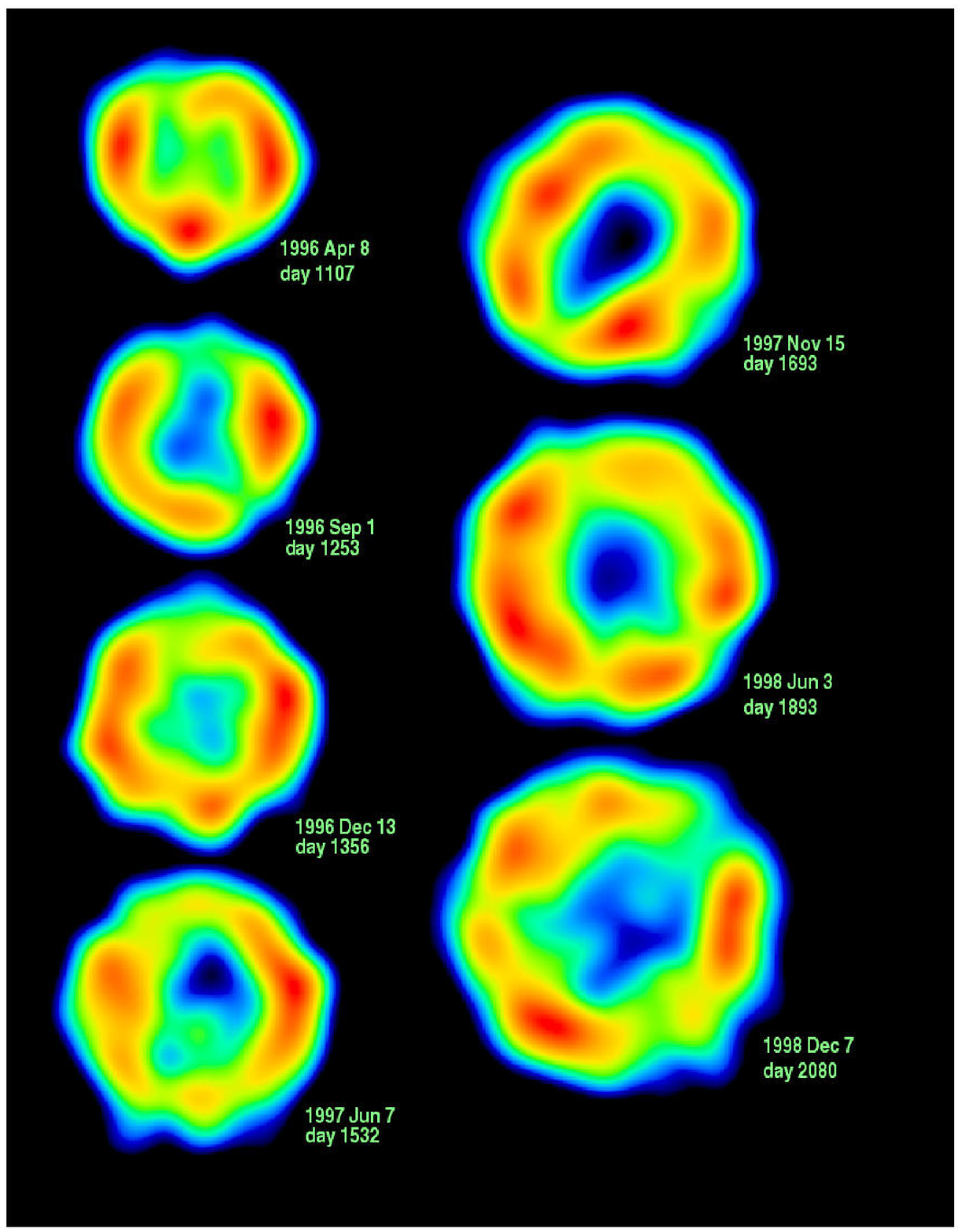}
\end{figure}

\clearpage 

\end{document}